\documentclass{aa}  

\usepackage{graphicx}
\usepackage{txfonts}

\usepackage{lscape,longtable}
\usepackage{natbib}
\usepackage{amsfonts,amsmath,subfigure,ulem}
\usepackage{color,xcolor,colortbl}
\usepackage{xtab,booktabs}
\usepackage{hyperref}
\hypersetup{breaklinks,colorlinks,citecolor=blue,linkcolor=blue,urlcolor=red}
\bibpunct{(}{)}{;}{a}{}{,} 

\definecolor{Gray}{gray}{0.9}
\definecolor{LCyan}{rgb}{0.88,1,1}
\definecolor{maroon}{cmyk}{0,0.87,0.68,0.32}
\definecolor{forestgreen}{cmyk}{0.76,0,0.76,0.45}

\begin{document} 

\title{Study of solar brightness profiles in the $18$ -- $26$~GHz frequency range with INAF radio telescopes\\ II. Evidence for coronal emission}

%
\author{M.~Marongiu\thanks{marco.marongiu@inaf.it}\inst{1}
\and A.~Pellizzoni\inst{1}
\and S.~Righini\inst{2}
\and S.~Mulas\inst{1}
\and R.~Nesti\inst{3}
\and A.~Burtovoi\inst{4,3}
\and M.~Romoli\inst{4,3}
\and G.~Serra\inst{5}
\and G.~Valente\inst{5}
\and E.~Egron\inst{1}
\and G.~Murtas\inst{6}
\and M.~N.~Iacolina\inst{5}
\and A.~Melis~\inst{1}
\and S.~L.~Guglielmino\inst{7}
\and S.~Loru\inst{7}
\and P.~Zucca\inst{8}
\and A.~Zanichelli\inst{2}
\and M.~Bachetti\inst{1}
\and A.~Bemporad\inst{9}
\and F.~Buffa\inst{1}
\and R.~Concu\inst{1}
\and G.~L.~Deiana\inst{1}
\and C.~Karakotia\inst{2}
\and A.~Ladu\inst{1}
\and A.~Maccaferri\inst{2}
\and P.~Marongiu~\inst{1}
\and M.~Messerotti\inst{10,11}
\and A.~Navarrini~\inst{1,12}
\and A.~Orfei\inst{2}
\and P.~Ortu\inst{1}
\and M.~Pili\inst{1}
\and T.~Pisanu\inst{1}
\and G.~Pupillo\inst{2}
\and P.~Romano\inst{7}
\and A.~Saba\inst{5}
\and L.~Schirru\inst{1}
\and C.~Tiburzi\inst{1}
\and L.~Abbo\inst{9}
\and F.~Frassati\inst{9}
\and M.~Giarrusso\inst{4,7}
\and G.~Jerse\inst{10}
\and F.~Landini\inst{9}
\and M.~Pancrazzi\inst{9}
\and G.~Russano\inst{13}
\and C.~Sasso\inst{13}
\and R.~Susino\inst{9}
}

%
\institute{INAF - Cagliari Astronomical Observatory, Via della Scienza 5, I--09047 Selargius (CA), Italy                        
\and INAF - Institute of Radio Astronomy, Via Gobetti 101, I--40129 Bologna, Italy                                              
\and INAF - Astrophysical Observatory of Arcetri, Largo Enrico Fermi 5, I--50125 Firenze, Italy                                 
\and University of Florence, Department of Physics and Astronomy, Via Giovanni Sansone 1, I--50019 Sesto Fiorentino (FI), Italy 
\and ASI - c/o Cagliari Astronomical Observatory, Via della Scienza 5, I--09047 Selargius (CA), Italy                           
\and Los Alamos National Laboratory, Bikini Atoll Rd, Los Alamos, NM 87545, USA                                                 
\and INAF - Catania Astrophysical Observatory, Via Santa Sofia 78, I--95123 Catania, Italy                                      
\and ASTRON – The Netherlands Institute for Radio Astronomy, Oude Hoogeveensedijk 4, 7991 PD Dwingeloo, The Netherlands         
\and INAF - Turin Astrophysical Observatory, Via Osservatorio 20, I--10025 Pino Torinese (TO), Italy                            
\and INAF - Trieste Astronomical Observatory, Via Giambattista Tiepolo 11, I--34131 Trieste, Italy                              
\and Department of Physics, University of Trieste, Via Alfonso Valerio 2, I--34127 Trieste, Italy                               
\and NRAO - Central Development Laboratory, 1180 Boxwood Estate Rd, Charlottesville, VA 22903, USA                              
\and INAF - Astronomical Observatory of Capodimonte, Salita Moiariello 16, I--80131 Napoli, Italy                               
}

\date{Received November 28, 2023; accepted February 10, 2024}

\abstract
{
One of the most important objectives of solar physics is the physical understanding of the solar atmosphere, the structure of which is also described in terms of the density ($N$) and temperature ($T$) distributions of the atmospheric matter.
Several multi-frequency analyses show that the characteristics of these distributions are still debated, especially for the outer coronal emission.
}
%
{
We aim to constrain the $T$ and $N$ distributions of the solar atmosphere through observations in the centimetric radio domain.
We employ single-dish observations from two of the INAF radio telescopes at the K-band frequencies ($18$ -- $26$~GHz).
We investigate the origin of the significant brightness temperature ($T_B$) level that we detected up to the upper corona ($\sim 800$~Mm of altitude with respect to the photospheric solar surface).
}
%
{
To probe the physical origin of the atmospheric emission and to constrain instrumental biases, we reproduced the solar signal by convolving specific 2D antenna beam models.
The analysis of the solar atmosphere is performed by adopting a physical model that assumes the thermal bremsstrahlung as the emission mechanism, with specific $T$ and $N$ distributions.
The modelled $T_B$ profiles are compared with those observed by averaging solar maps obtained at $18.3$ and $25.8$~GHz during the minimum of solar activity (2018 -- 2020).
}
%
{
We probed possible discrepancies between the $T$ and $N$ distributions assumed from the model and those derived from our measurements.
The $T$ and $N$ distributions are compatible (within $25\%$ of uncertainty) with the model up to $\sim 60$~Mm 
and $\sim 100$~Mm 
of altitude, respectively.
}
%
{
The analysis of the role of the antenna beam pattern on our solar maps proves the physical nature of the atmospheric emission in our images up to the coronal tails seen in our $T_B$ profiles.
Our results suggest that the modelled $T$ and $N$ distributions are in good agreement (within $25\%$ of uncertainty) with our solar maps up to altitudes $\lesssim 100$~Mm.
The challenging analysis of the coronal radio emission at higher altitudes, together with the data from satellite instruments will require further multi-frequency measurements.
}


\titlerunning{Coronal emission}
\authorrunning{Marongiu et al.}

\maketitle

%

\section{Introduction}
\label{par:intro}

The comprehension of the physical processes which govern mass and energy flow in the solar atmosphere represents one of the fundamental goals of solar physics (for a review, see e.g. \citealp{Shibasaki11}).
The analysis of the physical properties which characterise the solar atmosphere (e.g., the density and the temperature) is crucial to better understand these processes.
Multi-frequency observations (from radio to EUV-X domain) of the Sun probe different layers of the solar atmosphere -- with respect to the photosphere level $R_{\odot,opt} = 695.66 \pm 0.14$~Mm (corresponding to $959.16 \pm 0.19$~arcsec at Sun-Earth distance of $1$~AU\footnote{In this paper, we reported the angular solar size normalised to $1$~AU in all our conversions from the "absolute" size (expressed in units of Mm), unless stated otherwise.}; \citealp{Mamajek15,Prsa16,Haberreiter08}) -- gaining insights into the thermal origin (free–free bremsstrahlung radiation) of the atmospheric emission of the quiet Sun.
Many atmospheric models are based on optical spectral analysis (e.g. \citealp{Vernazza73,Vernazza76,Vernazza81,Fontenla90,Fontenla91,Fontenla93,Fontenla02}), which allows us to describe the layers of the solar atmosphere in terms of the temperature and the density distributions.

At radio frequencies the quiet Sun is characterised by pure free-free emission at local thermal equilibrium (LTE).
Under LTE conditions, radio waves cannot propagate outwards through regions below a radius $R_{\omega}$ where the local plasma frequency $\omega_p$ equals the observing frequency.
$\omega_p$ is defined as $\sqrt{N_e e^2/(m_e \epsilon_0)}$, where $N_e$ is the electron number density, $e$ is the elementary charge, $m_e$ is the electron mass, and $\epsilon_0$ is the vacuum permittivity.
In this frequency domain, especially at MHz range, the atmosphere of the Sun (from the photosphere to the upper corona) -- and its structure -- were observed and modelled by several authors in the literature over many decades (e.g. \citealp{Newkirk61,Wild70,Dulk77,Kundu83,Wang87,Ramesh98,Krissinel05,Ramesh10,Zhang22}).
These models play a fundamental role in the interpretation of solar radio emission and in the study of outward-flowing coronal material into the solar wind \citep{Lallement86,Fludra99,MeyerVernet12}.
In addition, they have indicated variability in the corona, depending upon solar activity (e.g. \citealp{Saito77}).
Among these models, starting from the observations at $17$~GHz with the Nobeyama Radioheliograph (NoRH; \citealp{Nakajima94,Nakajima95}) \citet{Selhorst05,Selhorst08} developed a two-dimensional temperature and density model to reproduce the observations ranging from $1.4$ to $400$~GHz, assuming the thermal bremsstrahlung as the emission mechanism.
This theoretical model reproduces the brightness temperature of the quiet-Sun $T_{qS}$ at the disk centre, the centre-to-limb brightening distribution, the value of the solar radius $R_{\odot}$, and it provides an estimation of the local distribution of temperature and density of the solar atmosphere.
However, this model and other similar approaches (e.g. \citealp{Mercier15,Vocks18,Ramesh20}), are characterised for simplicity by a symmetrical solar disk.
This aspect cannot independently explain both the equatorial and the polar limb profiles, thus preventing an accurate three-dimensional description of the density and the temperature distributions of the atmospherical layers of the Sun.
The comparison between the theoretical and the observed data is crucial to improve the modelling of the solar corona, and hence to better understand the atmospheric structures in terms of density and temperature (e.g. \citealp{SaintHilaire12,Mercier15,Vocks18,Ramesh20}).

In this paper we focus on the study of the atmospheric emission through the analysis of the brightness profiles of about $300$ solar images, obtained through single-dish observations from the newly-appointed Medicina "Gavril Grueff" Radio Telescope (hereafter Grueff Radio Telescope) and the Sardinia Radio Telescope (SRT) at the radio K-band ($18.3$ -- $25.8$~GHz), during about half a solar cycle (from 2018 to mid-2023), in the frame of the SunDish project aimed at the development of solar imaging observing modes for the INAF radio telescopes \citep{Pellizzoni22}.
Through this project, accurate brightness temperature $T_B$ profiles of the solar disk were obtained and the solar size was deeply analysed, finding that the measured solar radii $R_{\odot}$ -- ranging between $959$ and $994$~arcsec -- calculated for each observing frequency and radio telescope, are consistent with the other values reported in the literature \citep{Marongiu24a}.
The coverage of the entire solar disk, the low noise (RMS of the solar maps $< 10$~K), and the low contribution of the absolute calibration in the uncertainty on $T_B$ ($\lesssim 3\%$), make our data set valuable for analysing and modeling the solar atmospheric emission in the radio K-band.
These aspects have allowed us to detect significant emission up to the weak outer corona contributing to the "tails" extending outside the solar disk in our brightness profiles.

We organise this paper as follows.
A brief review of the solar imaging observations with INAF radio telescopes is presented in Section~\ref{par:obs_datared}.
The description of techniques employed for analysing the solar atmosphere, and related results are reported in Section~\ref{par:data_an}.
In particular, to corroborate the physical origin of the weak features in our brightness profiles as seen in correspondence of the outer corona, we modelled our solar maps by accounting for a specific 2D model (Section~\ref{par:ant_beam}); the brightness temperature spectrum profiles of our observed maps are compared with the theoretical estimations -- discussed in the literature -- in Section~\ref{par:solar_radio_corona}.
After the discussion of our results in Section~\ref{par:disc_concl}, we give our conclusions in Section~\ref{par:concl_svil}.

\section{Observations and data reduction}
\label{par:obs_datared}

The data set used in this work was provided by single-dish observation campaigns at the central frequency ranging between $18.1$ and $26.1$~GHz, with the network of the INAF single-dish radio telescopes\footnote{\url{https://www.radiotelescopes.inaf.it}}.
This network includes the Grueff Radio Telescope, the Sardinia Radio Telescope (SRT), and the Noto\footnote{Solar observations setup and operations are not yet implemented at Noto.} Radio Telescope.
The duration of these solar campaigns covers over 5 years, from 2018 to mid-2023.
These solar observations are the core of the "SunDish Project" (PI: A. Pellizzoni) \footnote{\url{https://sites.google.com/inaf.it/sundish}}, in collaboration with INAF and ASI \citep{Pellizzoni19,Plainaki20,Pellizzoni22}, aimed at the imaging and monitoring of the solar atmosphere.

The 32-m Medicina Radio Telescope -- located near Bologna at 25~m elevation -- observes and monitors the Sun once a week since February 2018.
The solar observations are carried out through the K-band dual-feed receiver operating at central frequency ranging between $18.3$ and $25.8$~GHz.
The beam pattern of this receiver is characterised by a beam size ranging between $1.5$ and $2.1$~arcmin.
\begin{figure*} 
\centering
{\includegraphics[width=89mm]{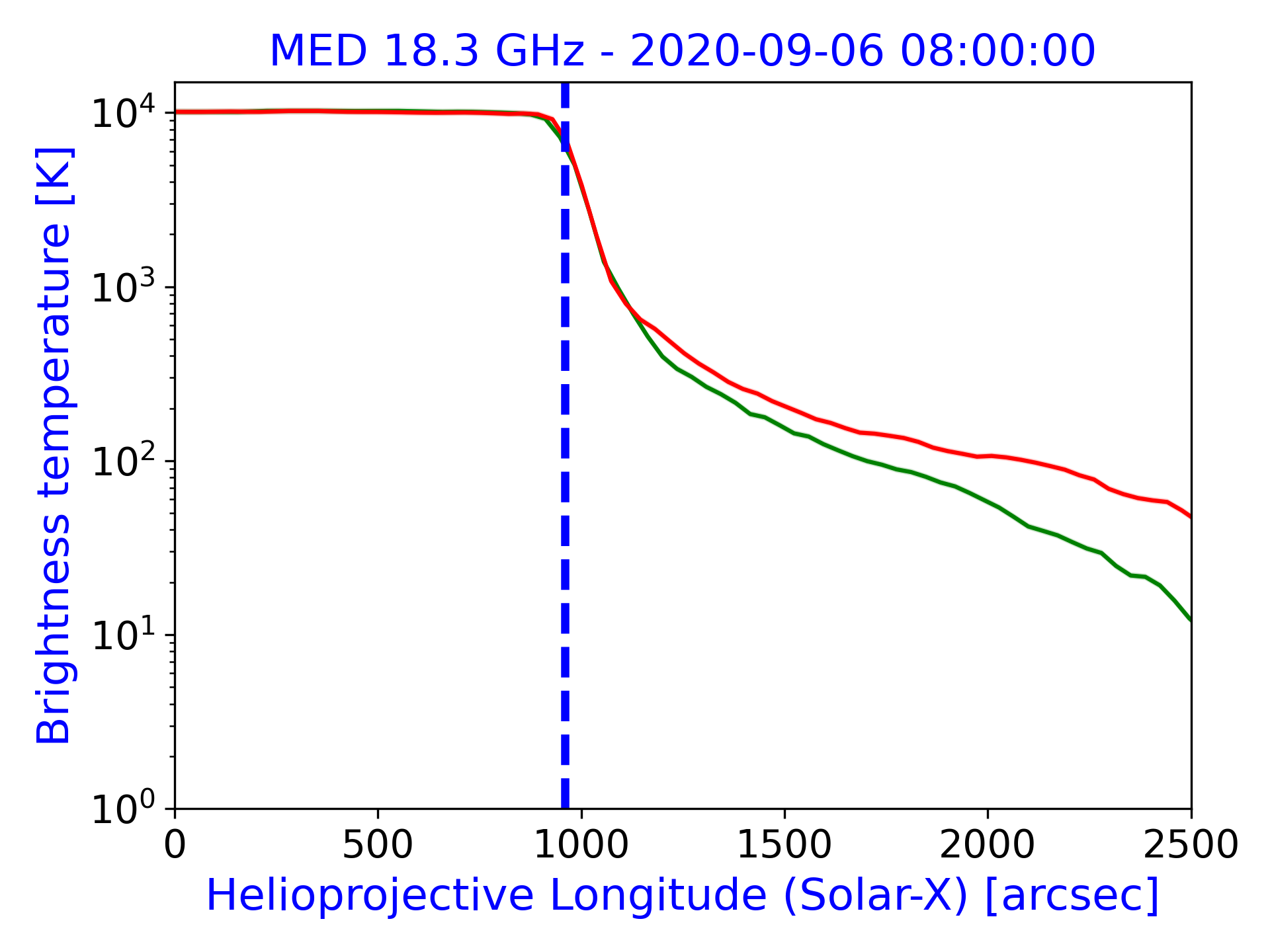}} \quad
{\includegraphics[width=89mm]{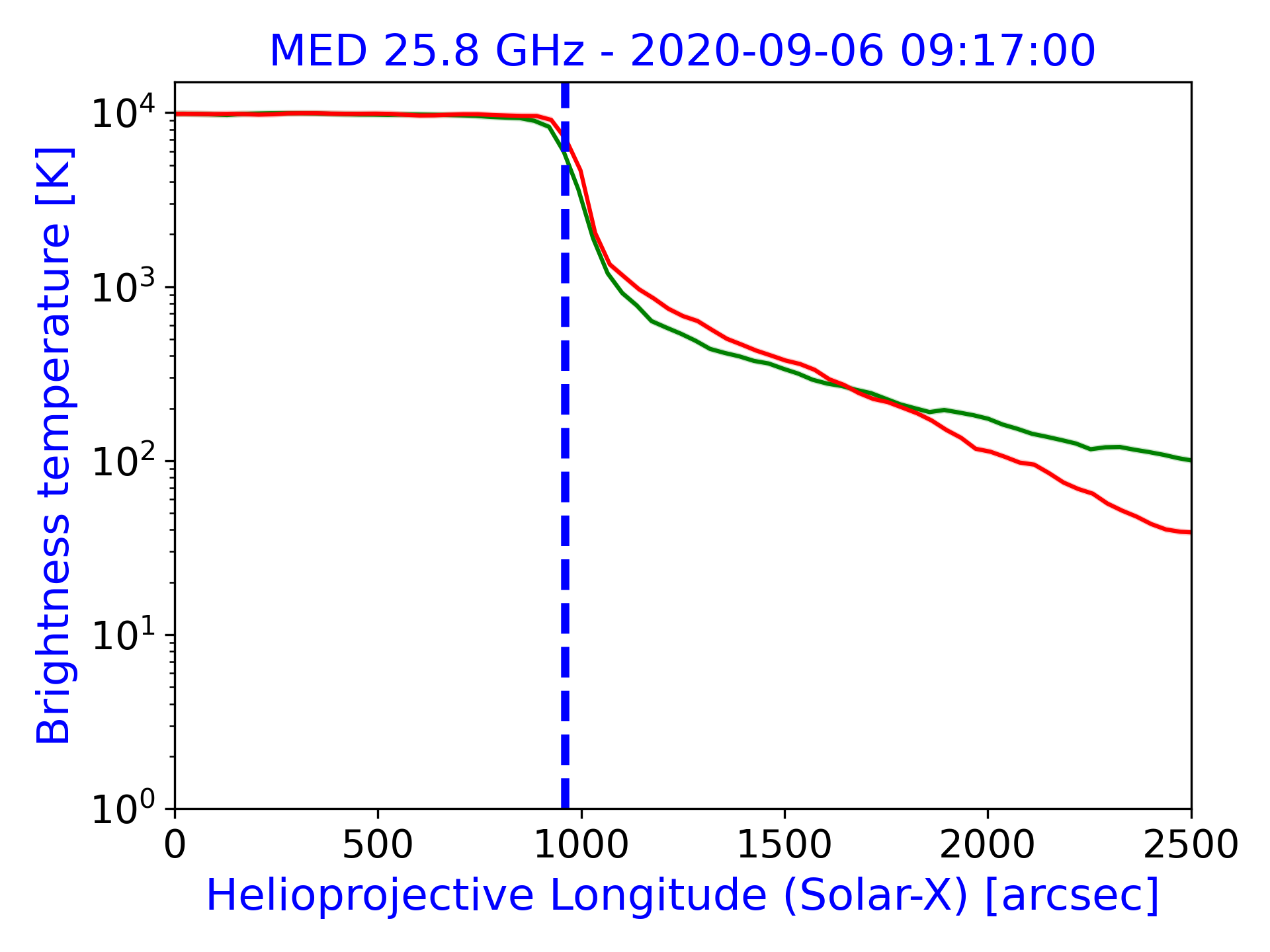}} \\
{\includegraphics[width=89mm]{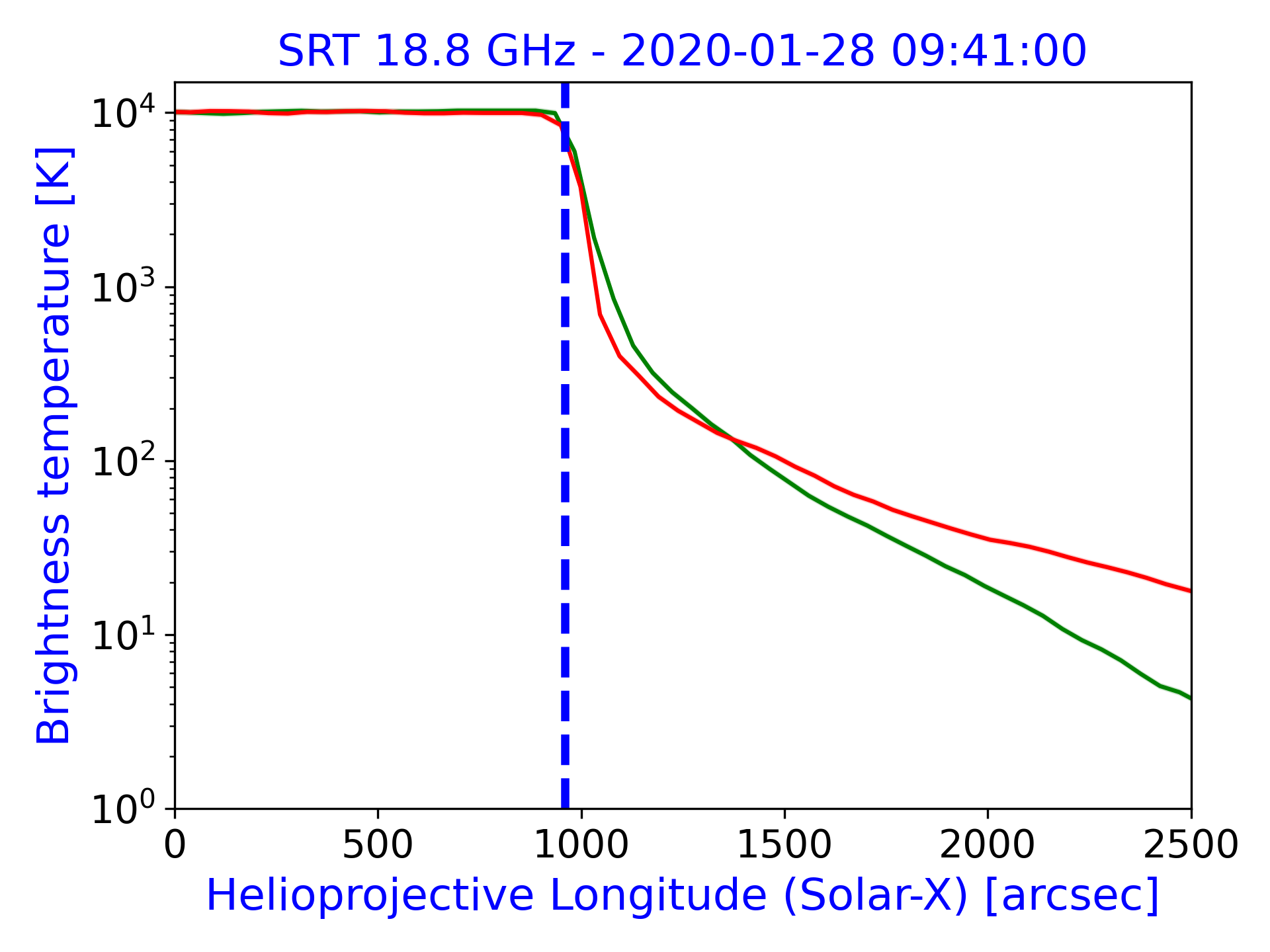}} \quad
{\includegraphics[width=89mm]{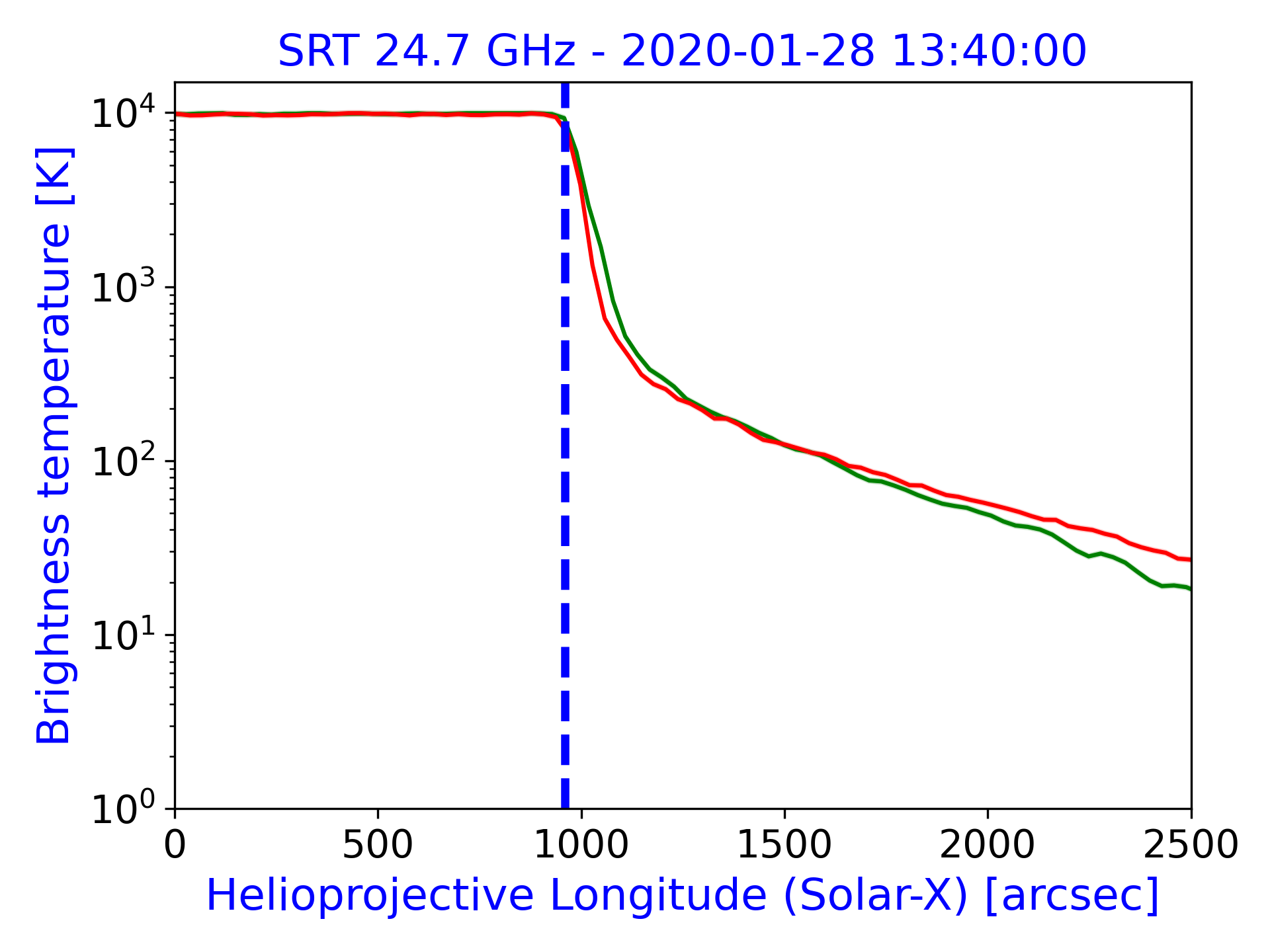}} \\
\caption{
Examples of polar (red line) and equatorial (green line) brightness temperature profiles derived from solar images obtained through the Grueff Radio Telescope and SRT.
(Top) Profiles derived on September 6, 2020 from the $18.3$~GHz (left) and $25.8$~GHz (right) Grueff Radio Telescope images.
(Bottom) Profiles derived on January 28, 2020 from the $18.8$~GHz (left) and $24.7$~GHz (right) SRT images.
Typical errors in the brightness profiles are in the $5$ -- $20$~K range, significantly affecting only the outwards tails.
Blue dashed lines indicate the solar radius at the photosphere level $R_{\odot,opt} = 959.16 \pm 0.19$~arcsec \citep{Mamajek15,Prsa16,Haberreiter08}.
}
\label{fig:sun_map_prof}
\end{figure*}
The 64-m SRT is located at $650$~m elevation in Sardinia (Italy).
Up to date, SRT has observed the Sun mostly at $18.8$ and $24.7$~GHz once a month through a $7$~feeds dual polarization K-band receiver, customised for solar observations \citep{Bolli15,Prandoni17,Pellizzoni22} and characterised by a beam size of $1.0$ and $0.8$~arcmin, respectively.
This radio telescope is currently in the final step of an upgrade phase that includes the installation of new receivers (suitable also for solar observations) operating up to $116$~GHz in the context of the National Operative Programme (Programma Operativo Nazionale-PON; \citealp{Govoni21})\footnote{\url{https://sites.google.com/a/inaf.it/pon-srt/home}}.

In the time frame $2018$ -- mid-$2023$, most of our solar maps in K-band were performed by the Grueff Radio Telescope.
For this work we employed an extensive data set of $290$ maps ($273$ maps from Grueff, and $17$ maps from SRT): in particular, this data set is composed of $145$ maps at $18.3$~GHz (Grueff), $128$ maps at $25.8$~GHz (Grueff), $10$ maps at $18.8$~GHz (SRT), and $7$ maps at $24.7$~GHz (SRT).
Mostly due to high atmospheric opacity or even instrumental errors, some maps had to be discarded ($\sim10\%$).

The radio signal is processed using the {\sc DISCOS} antenna control system\footnote{\url{http://discos.readthedocs.io/en/latest/user/index.html}} through full-stokes spectral-polarimetric ROACH2-based back-end (SARDARA system, \citealp{Melis18}) at SRT, and through Total-power/intensity back-end configuration at Medicina\footnote{From July 2021 this radio telescope is also equipped by a full-stokes spectral-polarimetric ROACH2-based back-end through the SARDARA system \citep{Mulas22}.
This back-end is available to the scientific community since the observing semester 2023A.}.
The solar maps are reconstructed from On-The-Fly (OTF) scans \citep{Prandoni17} in Right Ascension (RA) and Declination (Dec), respectively, covering an area in the sky of $4800 \times 4800$~arcsec at Medicina, and $5400 \times 5400$~arcsec at SRT.
The imaging and calibration procedures are performed using the SRT Single-Dish Imager (SDI; \citealp{Egron17,Pellizzoni19,Pellizzoni22,Loru19,Loru21,Marongiu20,Marongiu22,Mulas22}).
The data analysis is provided by the Python solar pipeline SUNPIT \citep{Marongiu22}.
Flux density and brightness temperature calibrations are achieved through the exploitation of the Supernova Remnant Cas~A (for details see \citealp{Pellizzoni22, Mulas22}).
Both for the Grueff Radio Telescope and SRT, details about the mapping techniques, the setup configurations, the observing strategy, and the data processing are available in \citet{Pellizzoni22}.
The data set of Grueff and SRT is available in the SunDish Archive\footnote{\url{https://sites.google.com/inaf.it/sundish/sundish-images-archive/sundish-archive-summary}}.

\section{Data analysis}
\label{par:data_an}

In this work, we analyse the properties of the solar atmospheric layers in terms of the temperature and the density distributions also above the solar radius altitude.
This kind of analysis was originated from the presence of "tails" extending outside the solar disk signal in our brightness profiles, as shown in Fig.~\ref{fig:sun_map_prof}.
These tails are characterised by the following features:
\begin{itemize}
    \item no significant differences are observed between polar and equatorial brightness temperature $T_B$ within errors (ranging between $10$ and $20$~K), as well as between Grueff and SRT measurements;
    \item there are no significant fluctuations of $T_B$ over time for each observing frequency, apart for statistical fluctuations compatible with noise;
    \item the $T_B$ level of the tails at high K-band ($\sim 25$~GHz) is higher than the $T_B$ level at low K-band ($\sim 18$~GHz), suggesting a thermal nature of the tail (coronal) emission (we estimate a spectral index\footnote{For this estimation we considered a preliminary coronal ring defined between $1.1$ and $1.7 R_{\odot}$.} $\alpha \gtrsim 1.5$);
    \item there is no correlation between the tails and the elevation $\delta$ of the Sun during the observations both at SRT and Grueff radio telescopes ($\delta$ ranges between $20$ and $60$ degrees).
\end{itemize}

To probe the physical origin of these tails -- connected with the atmospheric emission -- in our solar images, and to obtain more robust measurements of the solar radii \citep{Marongiu24a}, we analysed the degradation effects on the solar images due to the convolution of the antenna beam pattern with the solar signal (Sect.~\ref{par:ant_beam}).
Moreover, we compared the brightness temperature profiles of our observed solar maps with those obtained through a specific atmospheric model, assuming the thermal bremsstrahlung as the main emission mechanism (Sect.~\ref{par:solar_radio_corona}).

For this work we used $290$ solar maps ($273$ with Grueff and $17$ with SRT) to derive the brightness profiles and analyse the atmospheric emission in the range $18$ -- $26$~GHz.
We selected four frequencies, characterised by a uniform epoch coverage from 2018 to date: $18.3$ and $25.8$~GHz for Grueff, $18.8$ and $24.7$~GHz for SRT.
Moreover, we also used two averaged solar maps at $18.3$ and $25.8$~GHz obtained with the Grueff Radio Telescope during the minimum solar activity ($2018$--$2020$).

\subsection{Prescription to probe the role of the antenna beam pattern in the observed atmospheric emission}
\label{par:ant_beam}

One of the most important features that influence solar radio measurements is the degrading effect of the antenna beam pattern on the solar signal (e.g. \citealp{GimenezDeCastro20,Menezes22}).
As we will see, the analysis about the role of the antenna beam pattern that we have carried out to evaluate its impact on the solar radii measurements and to assess the radius determination (see \citealp{Marongiu24a}), also allows us to probe the physical nature of the external coronal emission observed as "tails" in the brightness profiles of our images of the Sun (see Fig.~\ref{fig:sun_map_prof}).

The Grueff and SRT beam patterns were reconstructed using a dedicated software called GRASP\footnote{\url{https://www.ticra.com/software/grasp/}}; except the measurement noise and some residual optical aberration, the beam patterns measured with the Grueff and SRT radio telescopes \citep{Prandoni17,Egron22} are in good agreement with the simulated ones obtained with GRASP.
We used the 2D beam patterns obtained through four observing frequencies ($18.3$ and $25.8$~GHz for Grueff; $18.8$ and $24.7$~GHz for SRT), the most observed by the Grueff and SRT radio telescopes during solar observations.
The peaks of the main and secondary lobes of these beam patterns are reported in Table~\ref{tab:beam_lobes}.
Adopting the GRASP beam model, the secondary lobes cannot account for the observed outer tails in the brightness profiles, as an artifact due to instrumental bias.
In fact, beam lobes can drive emission from the solar disk to the tail at a level only $<100$ -- $1000$ times lower (depending on the observing frequency) than the disk emission itself, while the tail brightness is still $>$1\% of that at $>8$ -- $10$~arcmin from the disk limb (Fig.~\ref{fig:sun_map_prof}).
Thus, our solar maps show a signal from the high solar atmosphere that is incompatible with instrumental biases caused by the degrading effect of the antenna beam pattern on the solar signal.

\begin{table}
\caption{
Values of the amplitudes of the main ($a_0$) and secondary ($a_2$) lobes (in units of dBi) obtained with GRASP for the analysed beam patterns of Grueff and SRT radio telescopes.
$\Delta a$ reports the difference between $a_2$ and $a_0$, and $R_{off}$ indicates the offset (in units of arcmin) between the secondary and the main lobes centroids.
}
\label{tab:beam_lobes}
\centering
\begin{tabular}{cc|cccc}
\hline
\hline
Frequency & Telescope & $a_0$  & $a_2$ & $\Delta a$ & $R_{off}$ \\
(GHz)     &           & (dBi)  & (dBi) & (dBi)      & (arcmin)  \\
\hline
18.3      & Grueff  & $73.5$ & $53.0$ & $-20.5$   & $3.1$     \\
25.8      & Grueff  & $76.2$ & $48.0$ & $-28.2$   & $2.7$     \\
\hline
18.8      & SRT       & $80.4$ & $61.4$ & $-19.0$   & $1.5$     \\
24.7      & SRT       & $82.9$ & $60.8$ & $-22.1$   & $1.3$     \\
\hline
\end{tabular}
\end{table}

For the sake of completeness, to properly take into account possible instrumental biases in our images and the effects of hypothetical instabilities of the opto-mechanical performances of the radiotelescopes during solar observations, we provided a parametrized 2D beam model derived from the original obtained by GRASP, along the lines of other similar works (e.g. \citealp{Iwai17}).
We modelled the generic beam pattern as a function characterised by (1) the instrumental (fixed) amplitude $a_0$ of the main lobe (Table~\ref{tab:beam_lobes}), and (2) a variable amplitude $A_{suc} = \mathcal{A} \times a_{suc}$ of the successive lobe peaks (Table~\ref{tab:priors}), where $a_{suc}$ is the instrumental (fixed) amplitude of the successive lobes, and $\mathcal{A}$ is the degradation level of the beam pattern (free parameter).
We define $\mathcal{A}$ as a dimensionless constant that describes the amplitude of the secondary lobe, accounting for possible unexpected optical deformations and aberrations.
The amplitude of the lobes in this model is expressed in units of dBi, and it adheres to the principle of energy conservation, maintaining constant the volume of the generic beam pattern.
In this context, $\mathcal{A}$ is taken inversely proportional to the width of the main lobe.

We convolved this parametrized beam model with specific solar signal models (see below) in order to fit our observed solar images.
This procedure (hereafter named "beam fitting procedure") provides as output the best-fit parameters for (1) the generic beam model adopted as input ($\mathcal{A}$, and hence the modelled secondary lobe amplitude $A_2$), and (2) the $R_{\odot}$ values of the adopted solar signal model.
The best-fit parameters of the beam fitting procedure are obtained thanks to a Bayesian approach based on Markov Chain MonteCarlo (MCMC) simulations (e.g. \citealp{Sharma17}).
For this approach we used the Python {\fontfamily{pcr}\selectfont emcee} package\footnote{\url{https://emcee.readthedocs.io/en/stable/}} \citep{Foreman13}.
The complete uniform prior distribution adopted in our analysis is listed in Table \ref{tab:priors}, where the parameters labelled with 0 indicate the initial best-fit parameters.
All the uncertainties are reported at 68\% ($1 \sigma$).
Further details about this approach applied to 2D solar signal models are available in \citet{Marongiu24a}.

\begin{figure} 
\centering
{\includegraphics[width=90mm]{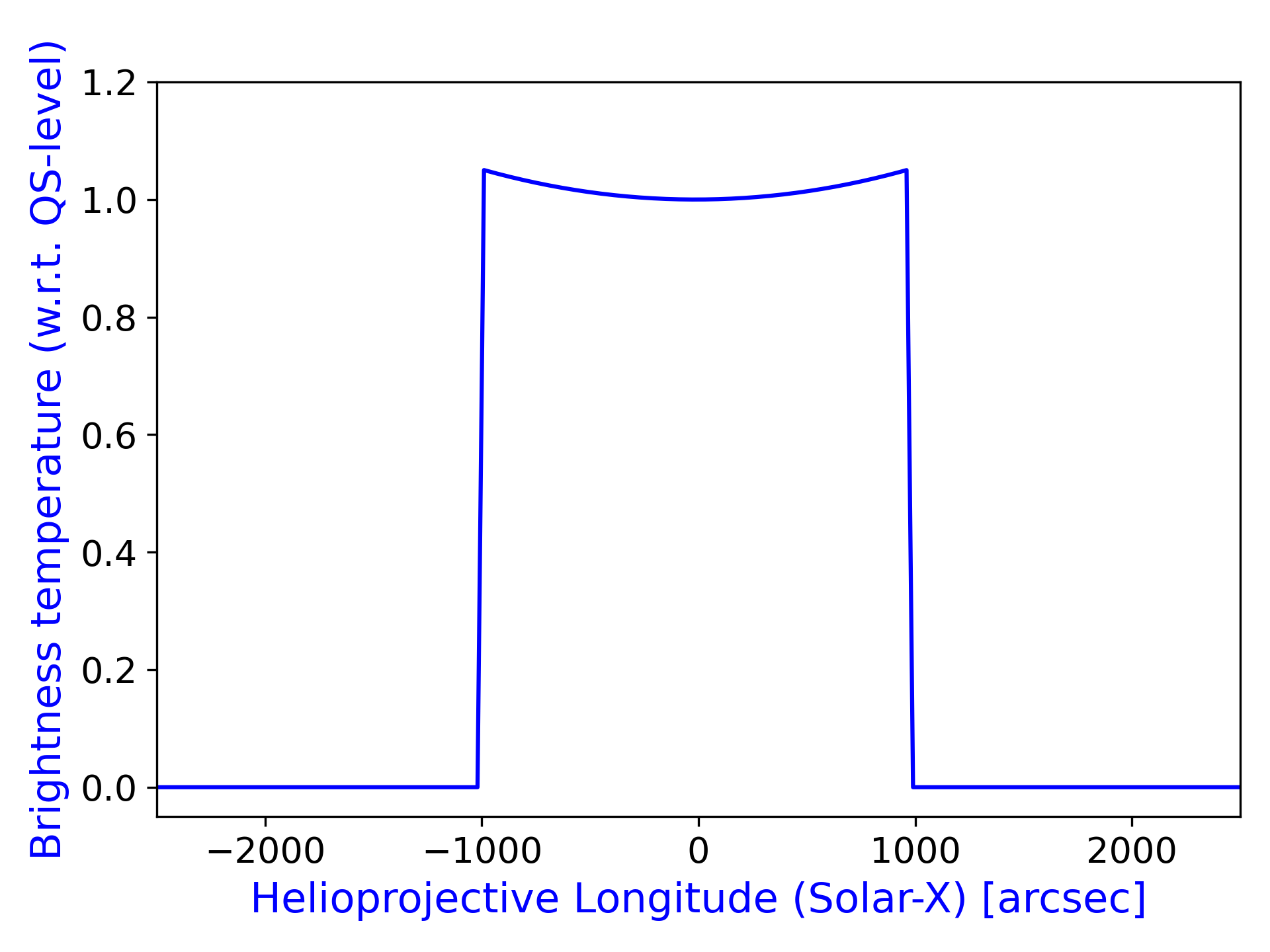}}
\caption{
Example of modelled equatorial $T_B$ profile of a solar map obtained adopting the CCBH-model as the solar signal.
}
\label{fig:ccbh_mod_example}
\end{figure}

The level of the "tails" seen outside the solar disk in the brightness profiles of the 2D-convolved solar maps is directly proportional to $\mathcal{A}$.
\begin{figure} 
\centering
{\includegraphics[width=88mm]{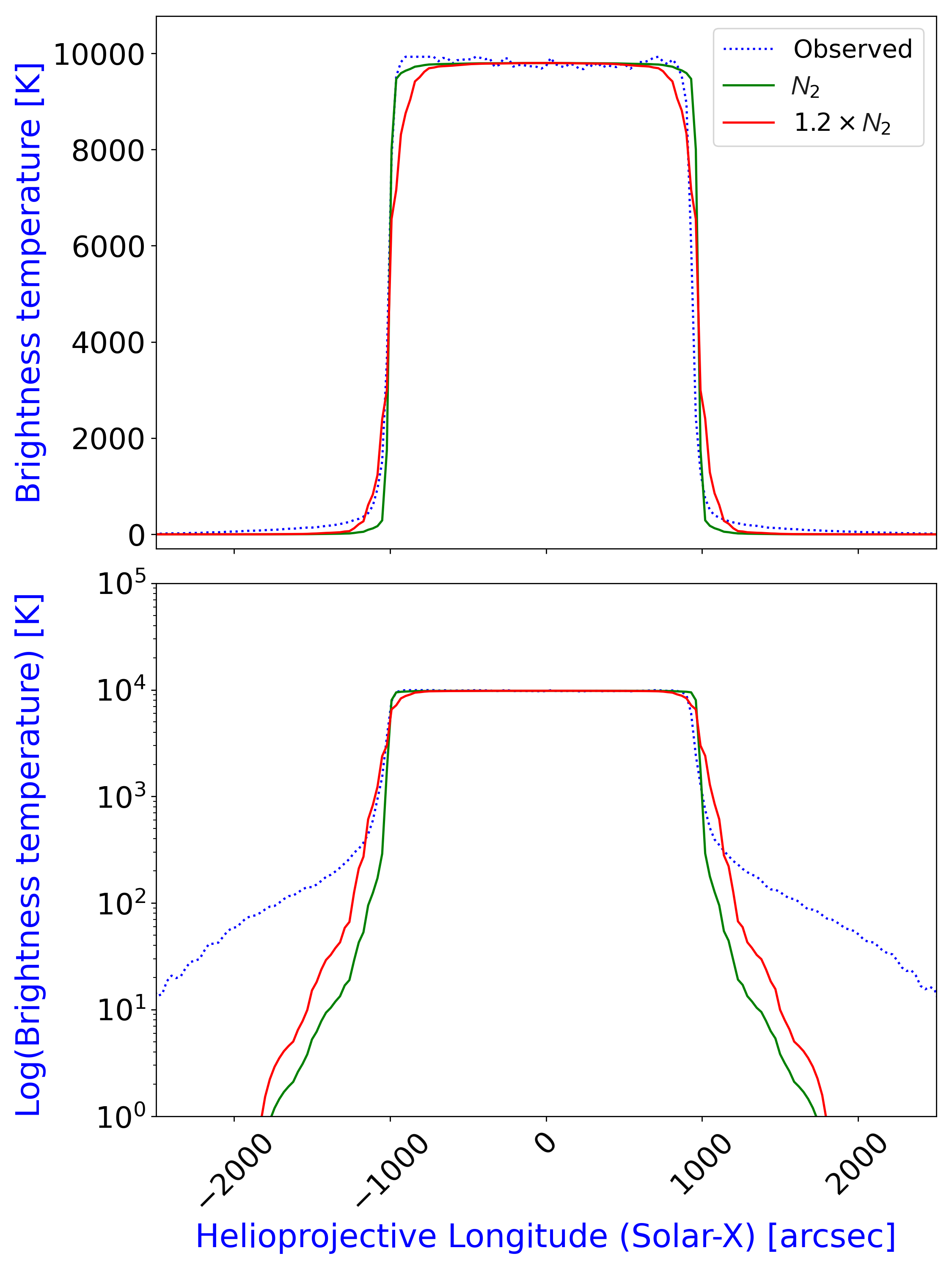}}
\caption{
Example of modelled equatorial $T_B$ profiles of a solar map obtained adopting the ECB-model as the solar signal for SRT at $24.7$~GHz (October 9, 2019), in comparison with the observed brightness profile.
These plots differ in the $T_B$ y-axis scale (linear on top, logarithmic on bottom). 
Blue dotted line indicates the observed equatorial $T_B$ profile of a solar map at this observing frequency with SRT, green solid line indicates the equatorial $T_B$ profile obtained convolving the solar signal model with the GRASP original beam pattern, and red solid line indicates the modelled equatorial $T_B$ profile obtained with a generic beam pattern with the secondary lobe characterised by $A_2 = 1.2 \times a_2$ (where $a_2$ is the reference value obtained by GRASP).
}
\label{fig:beam_example}
\end{figure}
Adopting a simple Elliptical-based Cylindrical Box (ECB-model, described in \citealp{Marongiu24a})
that trivially describes a pure and flat disk emission as the solar signal, it is not possible to fit the observed brightness profile even in the assumption of significant beam degradation effects (Fig.~\ref{fig:beam_example}).
Furthermore, when the beam degradation level is significant (i.e. $\mathcal{A} \gtrsim 1.05$), the limb of the solar disk in the 2D-convolved map is more smoothed than the observed solar map, and hence the solar signal requires a horned structure in the limb of the solar disk to compensate for this limb smoothing in the 2D-convolved map.
Thus, we empirically introduce this feature through the combination between a Circular-based Cylindrical Box and a Horned function, called \textbf{CCBH-model} (Fig.~\ref{fig:ccbh_mod_example}), defined as:
\begin{equation}
    f_{M}(x,y,R,T_c,l,n) = \left\{
    \begin{array}{l l}
        U_M & {\rm if} \;  U_M <= T_c+l \\
        0 & {\rm otherwise}
    \end{array} \right.
    \label{eq:f_boxemme}
\end{equation}
where $R$ indicates the box radius, $T_c$ indicates the brightness temperature $T_B$ of the centre of the solar disk (quiet-Sun level), and $l$ is the limb height (with respect to $T_c$).
$U_M$ is the horned function, defined as:
\begin{equation}
    U_M(x,y,R,T_c,l) = T_c + | [b (U_C - x_C)^\zeta] |
    \label{eq:f_emme}
\end{equation}
where $b = l [(R - x_C)^{-\zeta}]$, and $\zeta = 2$ (fixed).
$U_C$ indicates the circular base of the cylindrical box, defined as:
\begin{equation}
    U_C(x,y) = \sqrt{(x - x_C)^2 + (y - y_C)^2}
    \label{eq:circle_base}
\end{equation}
where $x_C$ and $y_C$ indicate the centre coordinates of the circle (fixed at 0, as defined for the solar maps at SRT and Medicina).
The CCBH-model (Eq.~\ref{eq:f_boxemme} and Fig.~\ref{fig:ccbh_mod_example}) allows us to account for the behaviour of the signal observed at the solar limb, where generally the limb brightening and the coronal emission are observed and smoothed by the characteristics of the antenna beam pattern.

\begin{table}
\caption{Uniform prior distribution adopted for the Bayesian approach in the solar signal models of our analysis.
CCBH indicates the solar signal model, and GPB indicates the model of the generic beam pattern; these models are described in Sect~\ref{par:ant_beam}.
}
\label{tab:priors}
\begin{tabular}{l|ccc}
\hline
\hline
Model   & Parameter     & Range                                     & Unit   \\
\hline
GBP     & $\mathcal{A}$ & $0.1 \leq \mathcal{A} \leq 10$            & -      \\
\hline
CCBH    & $R$           & $R_{0} - 100 \leq R_{0} \leq R_{0} + 100$ & arcsec \\
CCBH    & $l$           & $0.5 \leq l_{0} \leq 10000$               & K      \\
\end{tabular}
\end{table}

To investigate possible instrumental biases affecting the "tails" of the solar disk, we tested the beam fitting procedure using both solar signal models (ECB-model; CCBH-model, Eq.~\ref{eq:f_boxemme}) to obtain the hypothetical beam pattern that would fit the data, and to check whether it is physically plausible.
In this context, we compared the modelled value of the amplitude $A_2$ of the secondary lobe with those obtained through GRASP ($a_2$, Table~\ref{tab:beam_lobes}), to find possible anomalies with respect to the standard GRASP beam model.
\begin{table*}
\caption{Values of the modelled secondary lobe $A_2$ of the hypothetical beam patterns needed to fit the observed tails in the range $18$ -- $26$~GHz with Grueff and SRT radio telescopes, in the assumption of ECB/CCBH solar models.
We report also the difference between the modelled and the real values $\Delta A = A_2 - a_2$, and the values of the mean, equatorial, and polar radii of the Sun ($R_c$, $R_{eq}$, and $R_{pol}$).
}
\label{tab:raggio_norm}
\centering
\begin{tabular}{lll|cccc|c}
\hline
\hline
Frequency & Instrument & Model &  $R_{c}$              & $R_{eq}$              & $R_{pol}$             & $A_2$                & $\Delta A$ \\
(GHz)     &            &       & (arcsec)              & (arcsec)              & (arcsec)              & (dBi)                & (dBi)      \\
\hline
18.3      & Grueff     & ECB   & -                     & $991.7^{+6.1}_{-2.9}$ & $987.6^{+3.1}_{-2.8}$ & $58.8^{+1.0}_{-0.8}$ & $5.8$      \\
18.3      & Grueff     & CCBH  & $982.7^{+2.0}_{-1.8}$ & -                     & -                     & $60.6^{+0.6}_{-0.6}$ & $7.6$      \\
\hline
18.8      & SRT        & ECB   & -                     & $980.3^{+2.7}_{-0.6}$ & $981.8^{+3.5}_{-0.6}$ & $65.5^{+0.2}_{-0.4}$ & $4.1$      \\
18.8      & SRT        & CCBH  & $979.3^{+0.7}_{-0.8}$ & -                     & -                     & $66.5^{+0.2}_{-0.3}$ & $5.1$      \\
\hline
24.7      & SRT        & ECB   & -                     & $978.7^{+0.5}_{-0.4}$ & $979.6^{+1.3}_{-1.8}$ & $67.0^{+0.3}_{-0.4}$ & $6.2$      \\
24.7      & SRT        & CCBH  & $977.6^{+1.4}_{-2.0}$ & -                     & -                     & $67.4^{+0.1}_{-0.3}$ & $6.6$      \\
\hline
25.8      & Grueff     & ECB   & -                     & $989.8^{+4.2}_{-2.9}$ & $987.4^{+1.3}_{-1.5}$ & $60.2^{+0.9}_{-1.0}$ & $12.2$     \\
25.8      & Grueff     & CCBH  & $977.1^{+1.5}_{-1.3}$ & -                     & -                     & $62.2^{+0.6}_{-0.4}$ & $14.2$     \\
\hline
\end{tabular}
\end{table*}
In Table~\ref{tab:raggio_norm} we report the intensity difference $\Delta A$ between the original GRASP lobe $a_2$ (reported in Table~\ref{tab:beam_lobes}) and the modelled lobe $A_2$ (using both the ECB- and CCBH-models) of the hypothetical beam patterns needed to fit the observed tails in our images.

In the observing frequency range $18$ -- $26$~GHz, the large values obtained for $\Delta A$ (ranging between $4.1$ and $14.2$~dBi) are incompatible with hypothetical opto-mechanical deformations both for Grueff and SRT radio telescopes.
In fact, mean beam aberration effects at the observed elevations are expected $\lesssim \, $3~dBi at worst \citep{Egron22}, thus suggesting a genuinely solar origin of the tail signals, rather than instrumental biases.

\subsection{The theoretical radio atmospheric emission from the quiet Sun}
\label{par:solar_radio_corona}

In addition to the empirical solar signal models described in Sect.~\ref{par:ant_beam}, we considered existing theoretical atmospheric models providing brightness profiles that can be compared with our observations after convolving them with our beam pattern functions.
In particular, we adopted the atmospheric SSC model \citep{Selhorst05,Selhorst19b}, which describes the full solar atmosphere from the photosphere to the coronal plasma up to an altitude $h_{SSC} = 39.3$~Mm above the solar surface.
For our analysis we tried to describe the full solar atmosphere up to an altitude $h_{ext} \sim 2000$~Mm above the solar surface, following a prescription similar to that used by \citet{Zhang22}.
Therefore, in this work we named this model "extended SSC model" (eSSC).
The altitude $h_{ext}$ corresponds to a distance from the centre of the Sun $D_{\odot} = 3.9$~$R_{\odot,opt}$ ($\sim 3700$~arcsec), suitable for analysing our solar maps, characterised by a size of about $5000 \times 5000$~arcsec (Sect.~\ref{par:obs_datared}).
This model does not take into account the effects of the strong magnetic fields in active regions, the spicules, the special features observed at the polar regions \citep{Selhorst19b}, and the geometry of radio wave refraction within the solar corona \citep{Smerd50}.
As clarified in \citet{Selhorst05}, the main purpose of this model is to reproduce the brightness temperature spectrum of the solar disk centre, in order to model the full quiet Sun atmosphere from the photosphere to the corona.

\begin{figure} 
\centering
{\includegraphics[width=91mm]{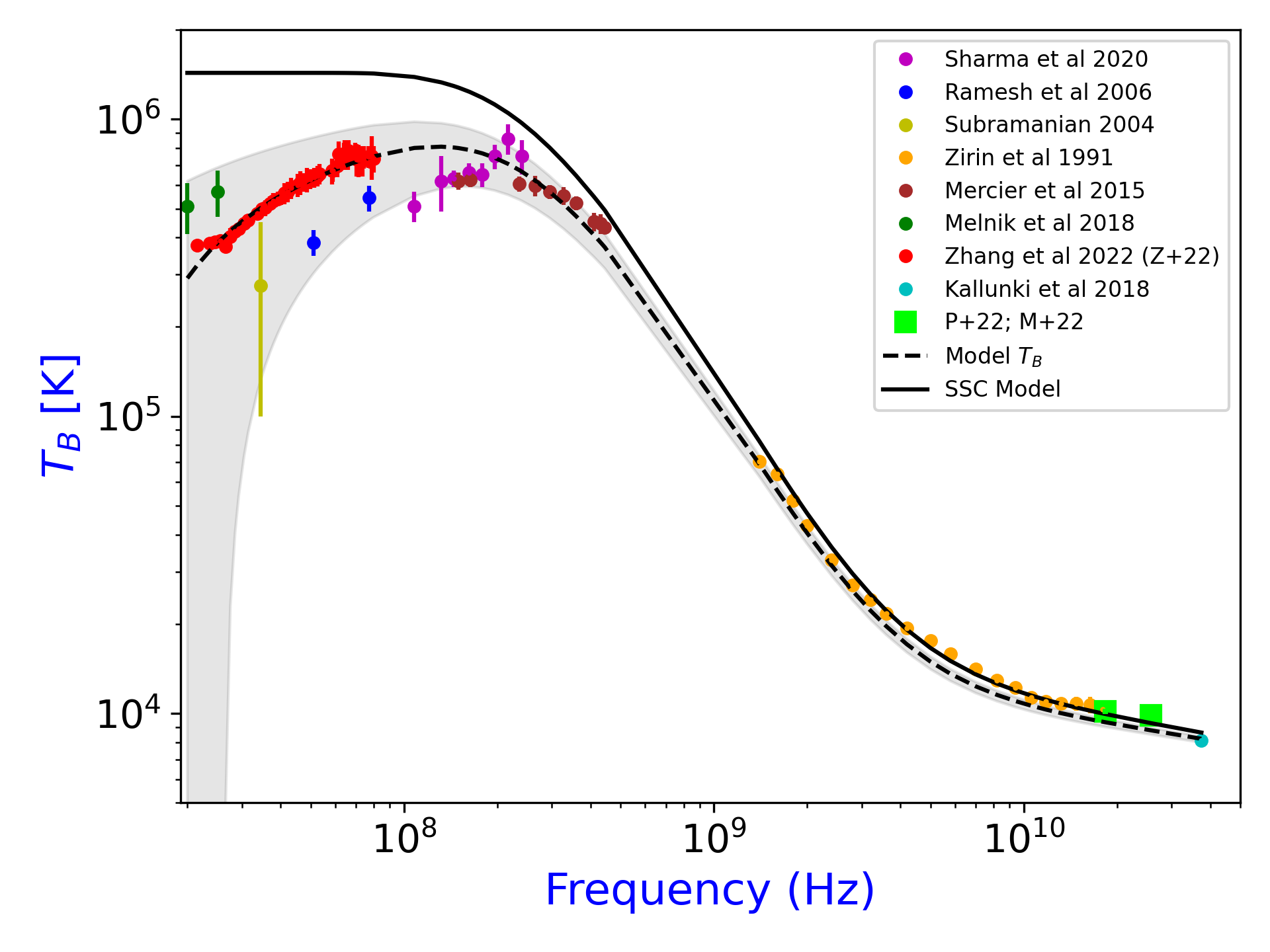}}
\caption{
Brightness temperature spectrum $T_{B} (\nu)$ for the solar disk centre obtained from the data of the INAF radio telescopes (averaged values $T_{B,av}$ from Medicina, Table~\ref{tab:overview_tb_mod}), and of previous works: \citet{Zirin91,Subramanian04,Ramesh06,Mercier15,Melnik18,Kallunki18,Sharma20,Zhang22}.
The black continuous line represents the modelled brightness temperature of bremsstrahlung emission obtained by the SSC model (\citealp{Selhorst05,Selhorst19b,Zhang22}); the black dashed line represents the fitted flux accounting for propagation effects; the dark grey shaded area shows the errors in the fit parameters.
Green points indicate the measurements obtained from the averaged solar maps of the Grueff Radio Telescope ($T_{B,av}$ in Table~\ref{tab:overview_tb_mod}), according to the analysis of \citet{Pellizzoni22} (P+22) and \citet{Mulas22} (M+22). 
This figure is an updated version of the figure reported by \citet{Zhang22}.
}
\label{fig:tb_spectra_example}
\end{figure}

In the eSSC model, $T_B$ -- expressed as a function of the observing frequency -- is derived from a model of the density and temperature of the solar atmosphere (basically composed of hydrogen) assuming the thermal bremsstrahlung as the emission mechanism, by numerically integrating the radiation transfer equation along the line of sight:
\begin{equation}
    T_{B} (\nu) = \int T k_{\nu} e^{-\tau_{\nu}} ds\;{\rm K}\,,
    \label{eq:t_b_model}
\end{equation}
where $T$ is the physical temperature of the medium (in units of K), $k_{\nu}$ is the free-free absorption coefficient (or the opacity, in units of $cm^{-1}$, \citealp{Zirin88,Simoes17}), and $\tau_{\nu} = \int k_{\nu} ds$ is the optical depth.
Eq.~\ref{eq:t_b_model} allows us to obtain theoretical brightness profiles since the information about the position with respect to the solar centre is contained in $ds$, thanks to simple geometrical transformations.
The $T$ distribution -- defined as a function of the altitude $h = R - R_{\odot}$ (in units of km) -- follows the SSC model up to $h_{SSC}$, with a constant extension at $T = 1.43 \times 10^6$~K (the last value of $T$ in the SSC model) for the outer corona (up to $h_{ext}$), as there is no rich information for the temperature of the outer corona in the literature.
Moreover, $k_{\nu}$ is expressed as
\begin{equation}
    k_{\nu} = 3.7 \times 10^8 T^{-1/2} N_e N_p \nu^{-3} g_{ff} (1-e^{- h \nu / k_b T})\;{\rm cm^{-1}}\,,
    \label{eq:kappa_nu}
\end{equation}
where $N_e$ and $N_p$ are, respectively, the density of electrons and protons in the solar atmosphere and outer corona, the term $(1-e^{- h \nu / k_b T})$ is the correction for stimulated emission, where $h$ and $k_b$ are the Planck and Boltzmann constants, respectively, and $g_{ff}$ is the Gaunt factor \citep{VanHoof14}.
The density distributions -- as a function of $h$ (in units of km) -- follow the SSC model up to $h_{SSC}$, with an extension for the outer corona (up to $h_{ext}$) as n times (with $n = 1.27$)\footnote{We used a different value of n with respect to that used in \citet{Zhang22} ($n = 1.25$) in order to have a perfect numerical connection between SSC and Saito models.} the density distribution described in a specific model \citep{Saito77}, in order to have a continuous connection with the SSC.
With the model described above, a brightness temperature spectrum $T_{B} (\nu)$ for the solar disk centre can be obtained by numerically integrating Eq.~\ref{eq:t_b_model} for a series of frequencies, as shown in Fig.~\ref{fig:tb_spectra_example}.
As reported by \citet{Selhorst05}, $T_{B} (\nu)$ calculated through the SSC model is in good accordance with the radio observations from $1.4$ to $400$~GHz.

Propagation effects (i.e., the refraction and scattering) were not included in the original SSC model.
These effects are very significant at lower frequencies and are inversely proportional to the observing frequency.
Thus we adopted a specific power-law term $T_B^{'}(\nu) = T_B(\nu)(1 - \alpha \nu^{\beta})$ -- previously used in \citet{Zhang22} -- to represent and model the attenuation of $T_B$ due to propagation effects.
In this term, $T_B^{'}(\nu)$ and $T_B(\nu)$ are, respectively, the observed and the modelled brightness temperature spectra, and $\nu$ is the observing frequency (in units of Hz).
The best-fit parameters for $\alpha$ and $\beta$ are reported in Table~\ref{tab:overview_tb_zhang} considering different data sets.
All the quiet-Sun measurements reported in Fig.~\ref{fig:tb_spectra_example} are well modelled (Tab.~\ref{tab:overview_tb_zhang}), but the best-fit parameters of our modelling ($\alpha$ and $\beta$) are different with respect the results obtained by \citet{Zhang22}, probably caused by a wider data set.
Moreover, our results indicate that the propagation effects are already negligible in the range $18$ -- $26$~GHz, and then these effects are not important for the modelling of the coronal tail emission.
As reported in Tab.~\ref{tab:overview_tb_zhang}, we pointed out that the parameters $\alpha$ and $\beta$ are highly sensitive to the variation of $T_B(\nu)$.
\begin{table}
\caption{Best-fit parameters obtained by the modelling of the power-law term that describes the attenuation of $T_B$ due to propagation effects.
}
\label{tab:overview_tb_zhang}
\scriptsize
\centering
\begin{tabular}{cc|ccc}
\hline
\hline
Data          & Model       & $\alpha$                      & $\beta$            & $\chi^2_r$ \\
\hline
LOFAR         & eSSC        & $(2.92 \pm 0.16) \times 10^2$ & $-(0.35 \pm 0.02)$ & $0.4$     \\
LOFAR/SunDish & eSSC        & $(3.87 \pm 0.20) \times 10^2$ & $-(0.37 \pm 0.02)$ & $0.6$     \\
All           & eSSC        & $(4.19 \pm 0.19) \times 10^2$ & $-(0.37 \pm 0.02)$ & $1.1$     \\
\hline
\end{tabular}
\end{table}

\begin{figure*} 
\centering
{\includegraphics[width=89mm]{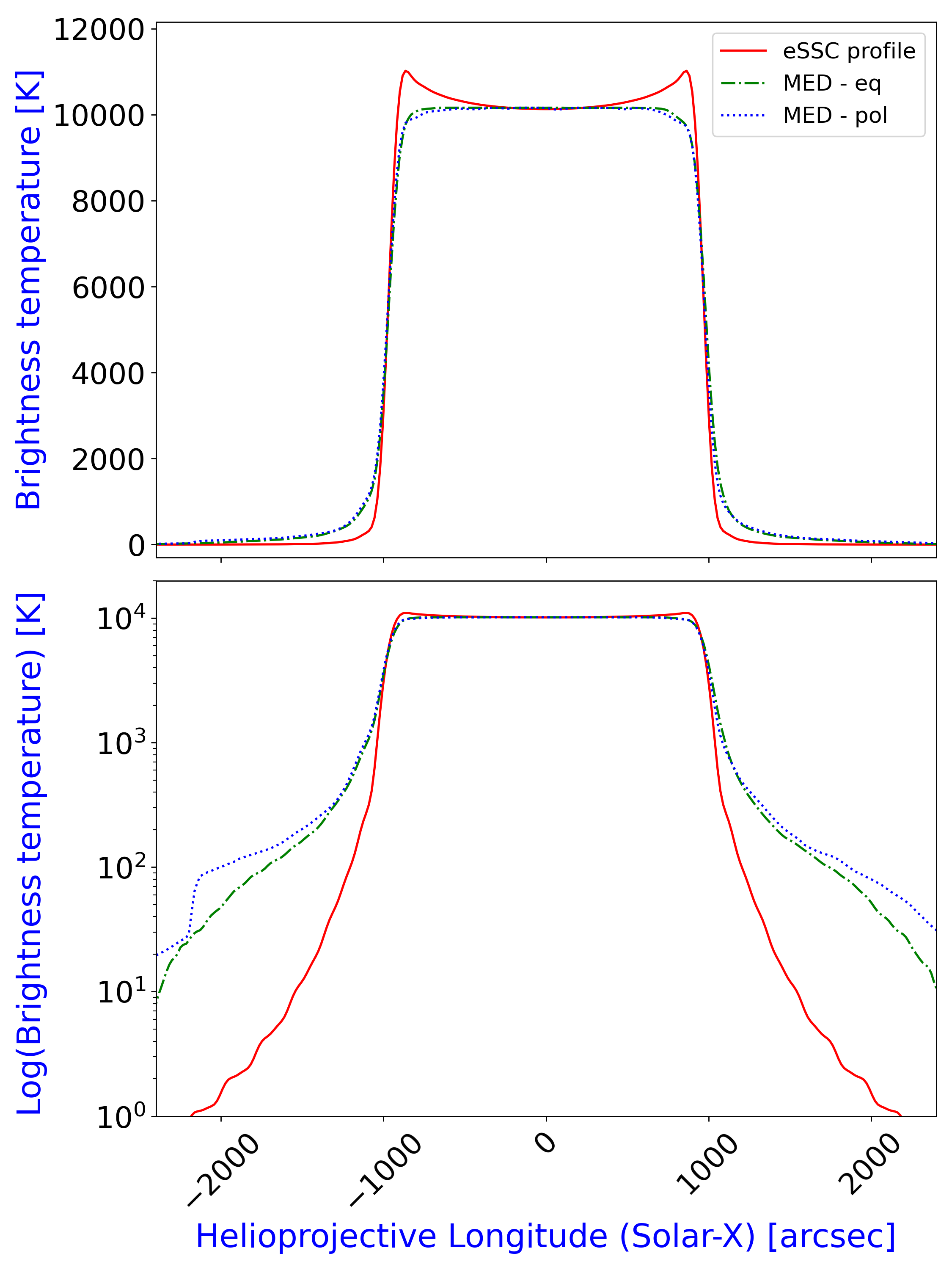}} \quad
{\includegraphics[width=89mm]{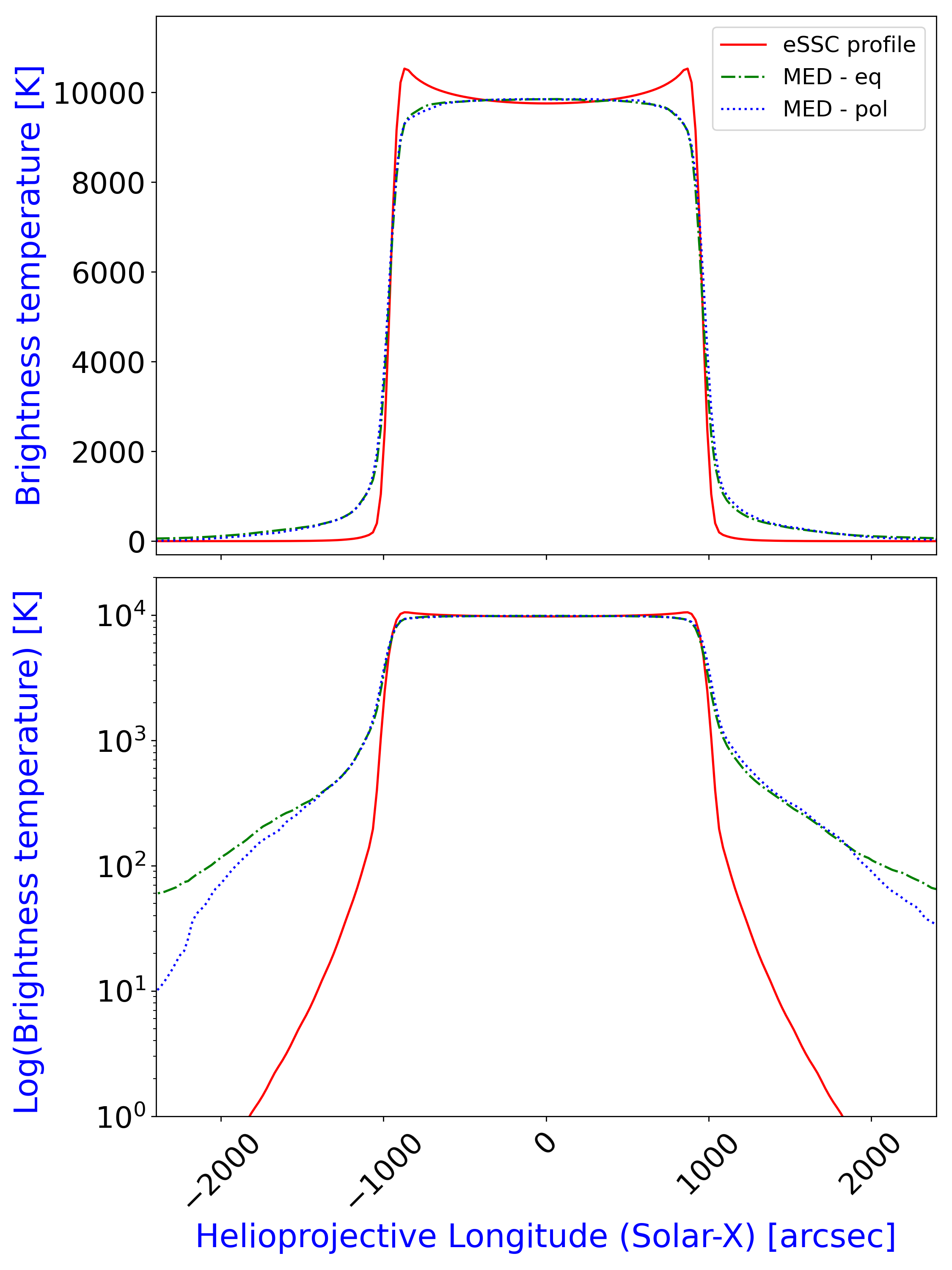}} \\
\caption{
Averaged and modelled $T_B$ profiles obtained at the observing frequencies of $18.3$ (left) and $25.8$ GHz (right) with the Grueff Radio Telescope.
These plots differ in the $T_B$ y-axis scale (linear on top, logarithmic on bottom).
Red solid line indicates the modelled $T_B$ profile, obtained from the convolution between the eSSC model (Eq.~\ref{eq:t_b_model}) and the beam pattern of the receiver for each observing frequency.
Green dot-dashed and blue dotted lines indicate the equatorial and polar profiles of the averaged solar map, respectively.
}
\label{fig:overview_maps_med}
\end{figure*}

To strengthen the evidence of the physical nature of the coronal emission in the tails of the $T_B$ profiles (Sect.~\ref{par:ant_beam}) and to constrain the temperature and density parameters of the atmospheric emission, we also compared the modelled and the observed $T_B$ profiles along the equatorial and polar diameters of the quiet Sun during the minimum solar activity ($2018$--$2020$), and their variation with the observing frequency.
The modelled $T_B$ profiles are obtained through the SSC and the eSSC model (Eq.~\ref{eq:t_b_model}), convolved with the beam pattern of the instrument for that observing frequency.
The observed $T_B$ profiles are obtained with the Grueff Radio Telescope through two averaged solar maps, one at $18.3$~GHz and the other at $25.8$~GHz (green dotted and blue dashed lines in Fig.~\ref{fig:overview_maps_med}).
These averaged solar maps have been created using only the medium/high-quality maps (to avoid systematic and/or meteorological undesired effects) during the minimum phase of the solar cycle, to minimise the effects of solar activity.
We selected these observing frequencies and this radio telescope because we have a moderately uniform time coverage of solar observations from 2018 to date.

From Fig.~\ref{fig:overview_maps_med} and Table~\ref{tab:overview_tb_mod}, we can compare the theoretical $T_B$ profiles with the observed ones using the averaged solar maps of the Grueff Radio Telescope.
In particular, we observe these features:
\begin{itemize}
    \item the level of the averaged tails of the $T_B$ profiles (shown as green dot-dashed and blue dotted lines for the equatorial and the polar profiles, respectively) is higher than those predicted by the eSSC model (shown as a red solid line) by a factor $4.5$ at $18.3$~GHz, and by a factor $13$ at $25.8$~GHz, suggesting that the difference between the model and observation increases with the increase in observing frequency;
    \item the strong limb brightening (about $10 \%$) predicted by the theory does not emerge from our averaged limbs in the $T_B$ profiles;
    \item the $T_B$ values -- at the central position of the solar disk -- obtained through the averaged solar maps are slightly higher than those obtained in the modelled $T_B$ profiles using the eSSC model, especially at $25.8$~GHz where the discrepancy is of $\sim 6\%$ (Table~\ref{tab:overview_tb_mod}).
\end{itemize}
The same behaviour in the comparison between the theoretical model and the averaged $T_B$ profiles occurs also when the SSC model is considered.

\begin{table}
\caption{Measured $T_B$ at the central position of the Sun -- obtained at $18.3$ and $25.8$~GHz with the Grueff Radio Telescope -- from (1) the modelled quiet-Sun level ($T_{B,qS}$) according to the Cas~A calibrator source \citep{Mulas22,Pellizzoni22}, (2) the averaged solar maps ($T_{B,av}$), and (3) the modelled value of the eSSC model, convolved with the beam pattern of the instrument ($T_{B,eSSC}$).
}
\label{tab:overview_tb_mod}
\centering
\begin{tabular}{c|ccc}
\hline
\hline
Frequency & $T_{B,qS}$      & $T_{B,av}$      & $T_{B,eSSC}$   \\
(GHz)     & (K)             & (K)             & (K)            \\
\hline
18.3      & $10130 \pm 253$ & $10159 \pm 254$ & $9957 \pm 100$ \\
25.8      & $9755 \pm 244$  & $9848 \pm 246$  & $9270 \pm 93$  \\
\hline
\end{tabular}
\end{table}

These features suggest that (1) the SSC/eSSC model could be too simplified to describe the radio emission from the quiet Sun, spanning from the photosphere to the upper corona (neglecting effects such as gyro-magnetic emission and propagation effects, as mentioned below in this Section), and/or (2) the $T$ and $N$ distributions of the SSC/eSSC model could differ from what we can derive from our observations.
In this context, we performed a modelling of our observed $T_B$ profiles (both equatorial and polar) for each observing frequency (Figs.~\ref{fig:profile_results_te}, \ref{fig:profile_results_n}, and \ref{fig:dens_sundish_metis}).
Our brightness profiles cannot univocally constrain the actual physical parameters required for the fit, but as a first approach we can check the effects of the variation of density and temperature separately.
A possible fitting solution for the $N_e$ and $N_p$ distributions can be obtained while maintaining constant the $T$ distribution according to the SSC/eSSC model, and, conversely, a valid $T$ distribution can be obtained while maintaining constant the original density distributions.
The best-fit results are obtained in iterative mode, as long as the value of $T$ (or $N$) allows us to model $T_{B} (\nu)$ of the Eq.~\ref{eq:t_b_model}, for each solar coordinate of the observed $T_B$ profiles.
The computational error connected with the calculation of the integral to obtain $T_{B} (\nu)$ (Eq.~\ref{eq:t_b_model}) is conservatively fixed at $1\%$.
In the fitting procedure we also carefully take into account the errors related to our brightness temperature measurements:
\begin{itemize}
    \item uncertainties due to the image RMS ($2.5$~K at $18.3$~GHz and $3.5$~K at $25.8$~GHz);
    \item errors ($2.5\%$) associated with the calibration procedure adopting the Supernova Remnant Cas~A as calibrator source \citep{Mulas22,Pellizzoni22};
    \item effects caused by the background baseline subtraction during the imaging procedure ($5$~K at $18.3$~GHz and $15$~K at $25.8$~GHz).
\end{itemize}

Accounting for these uncertainties we are able to analyse the solar atmosphere up to a specific $T_B$ threshold ($10$~K at $18.3$~GHz and $20$~K at $25.8$~GHz).
This implies for example that we can observe the corona with the Grueff Radio Telescope up to (1) $2400$~arcsec ($\sim 2.5\,R_{\odot}$) with respect to the Sun centre at $18.3$~GHz, and (2) $2600$~arcsec ($\sim 2.7\,R_{\odot}$) with respect to the Sun centre at $25.8$~GHz.
However, due to the high uncertainties starting from $\sim 1.8\,R_{\odot}$ (see Fig.~\ref{fig:dens_sundish_metis}), we consider reliable only the data up to $2050$~arcsec ($2.15\,R_{\odot}$).

\section{Discussion}
\label{par:disc_concl}

\begin{figure*} 
\centering
{\includegraphics[width=89mm]{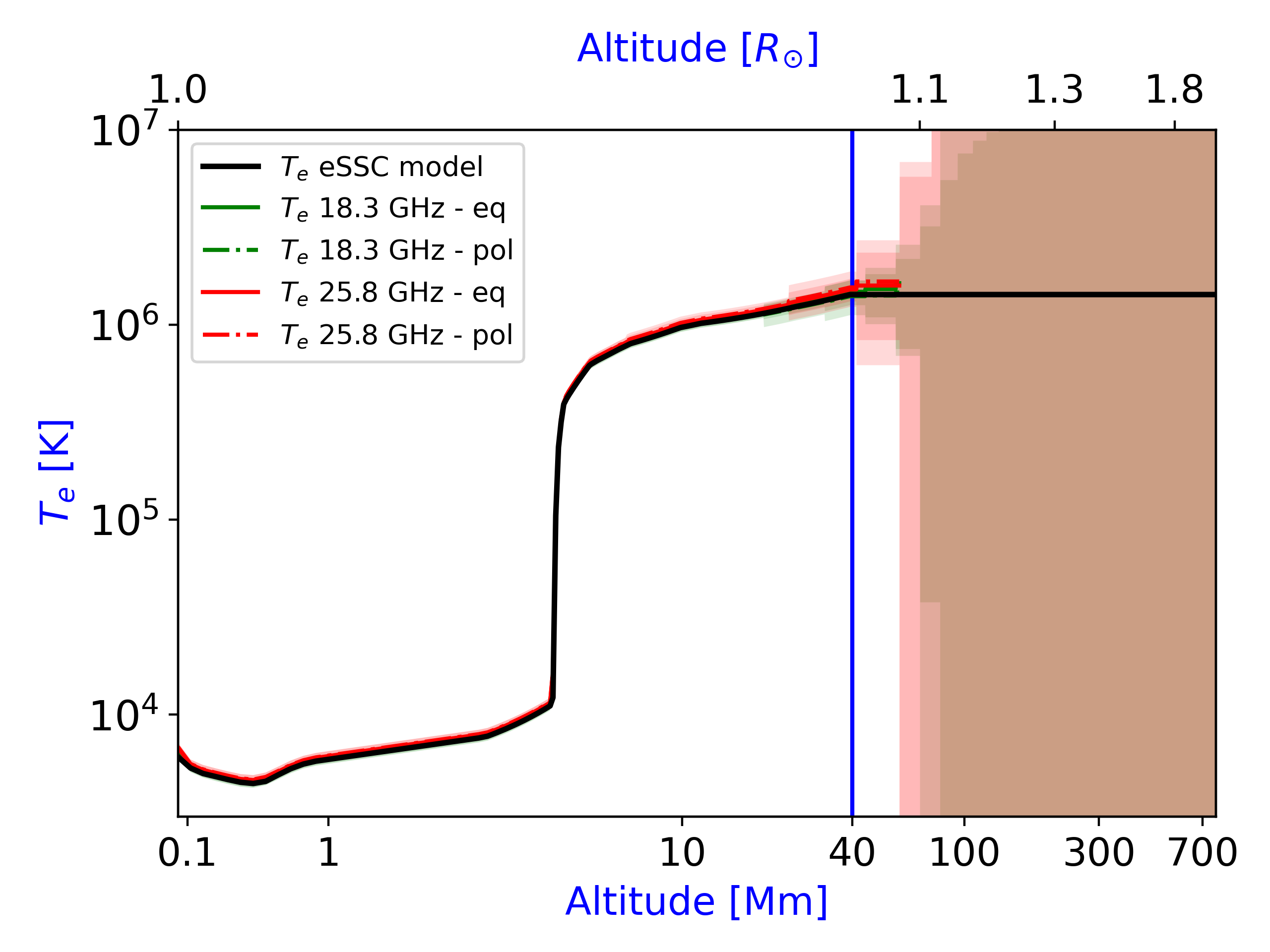}} \quad
{\includegraphics[width=89mm]{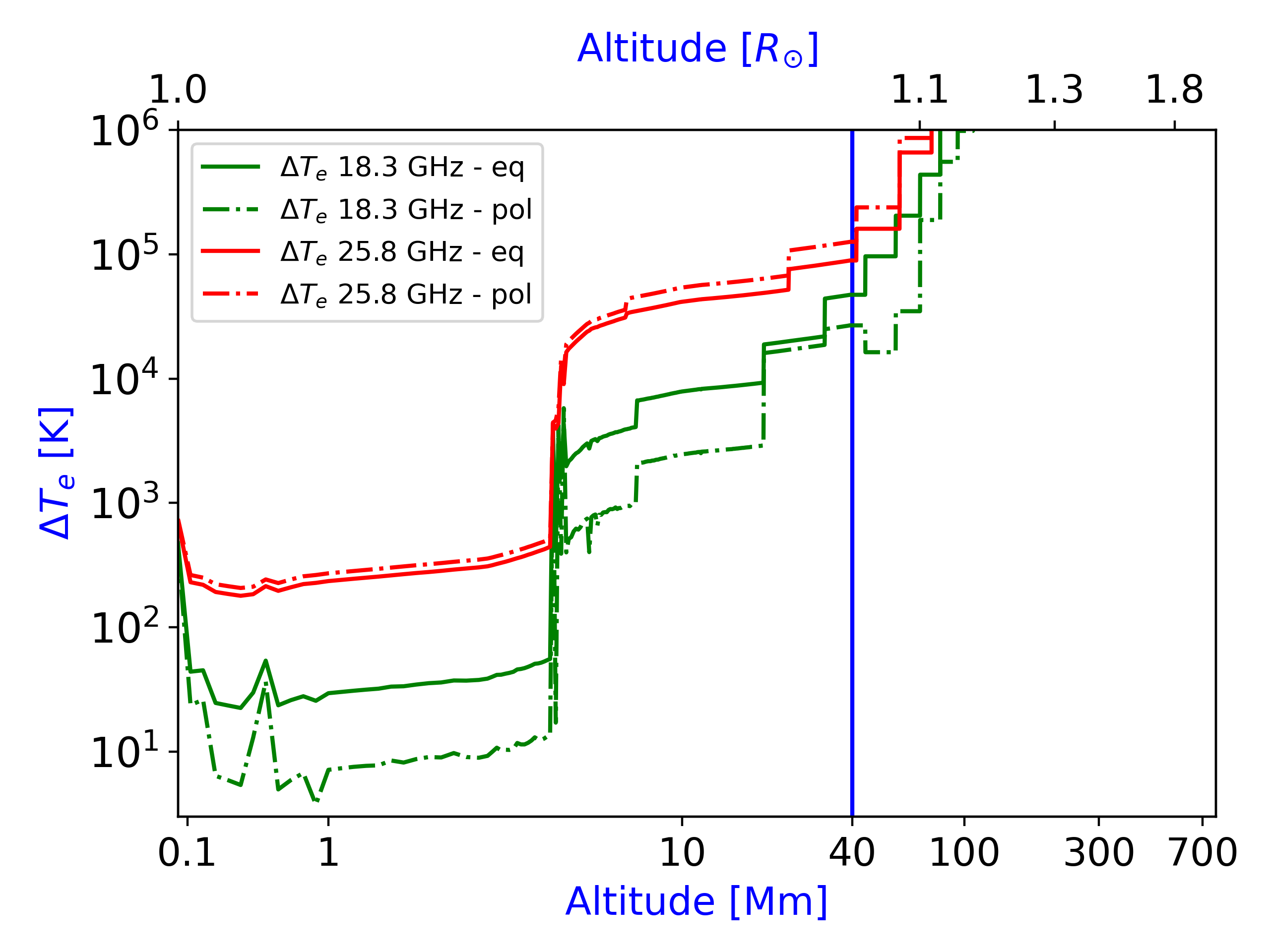}} \\
\caption{
Distribution of $T$ (left) and $\Delta T$ (right) as a function of the altitude from the solar surface, defined as the photosphere level $R_{\odot, opt}$.
The respective shaded area indicates the $1 \sigma$ error of these distributions.
The black line indicates the theoretical $T$ distribution of the original eSSC model.
The blue vertical line indicates the altitude $h_{SSC}$, below which the SSC model is defined.
The y-axis of these plots is shown in logarithmic scale.
Best-fit $T$ distribution is compatible (within an uncertainty $\lesssim 25\%$) with the original eSSC model below $\sim 100$~Mm; above this altitude, this distribution is characterised by strong degeneracy.
}
\label{fig:profile_results_te}
\end{figure*}

\begin{figure*} 
\centering
{\includegraphics[width=89mm]{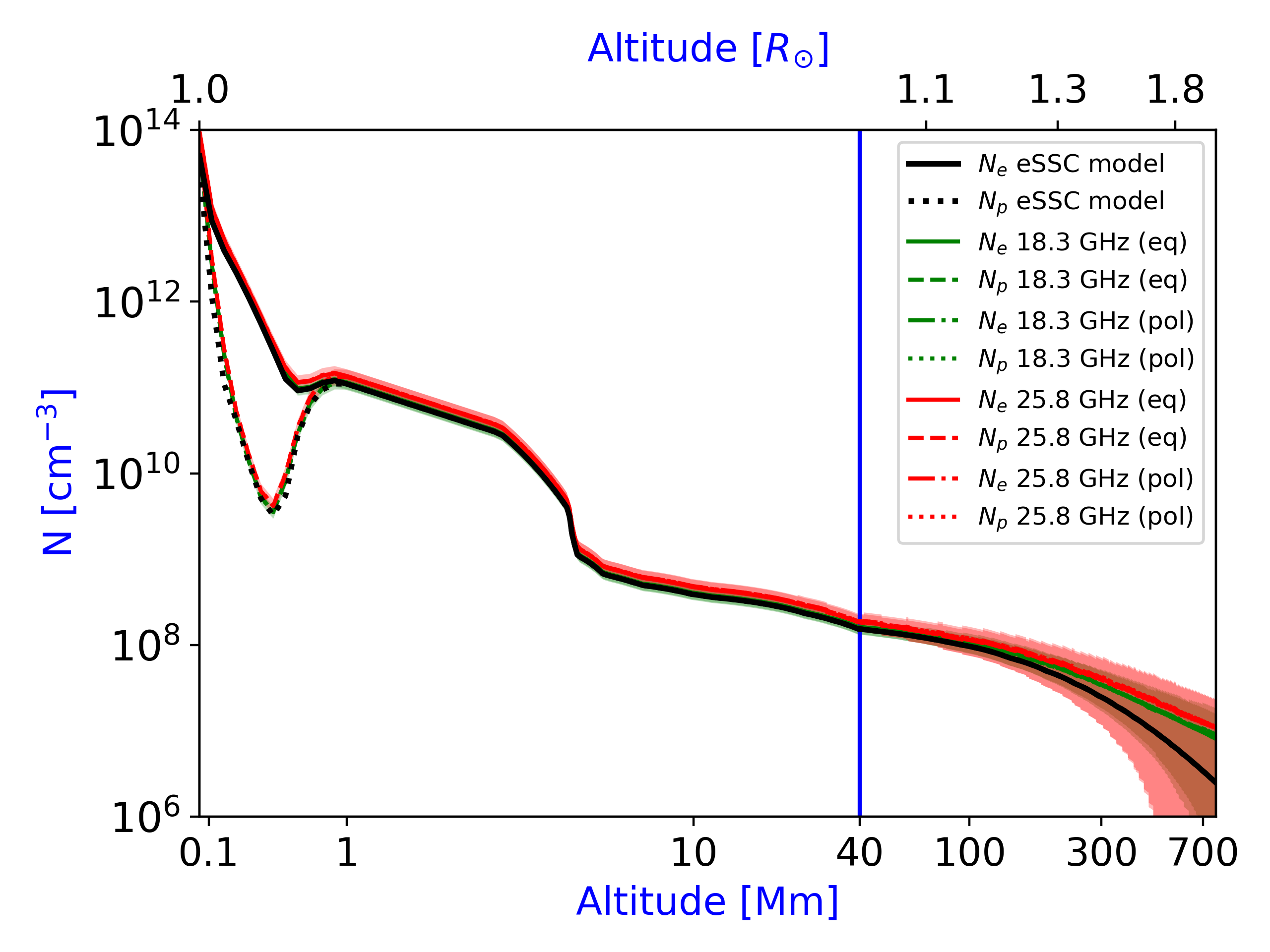}} \quad
{\includegraphics[width=89mm]{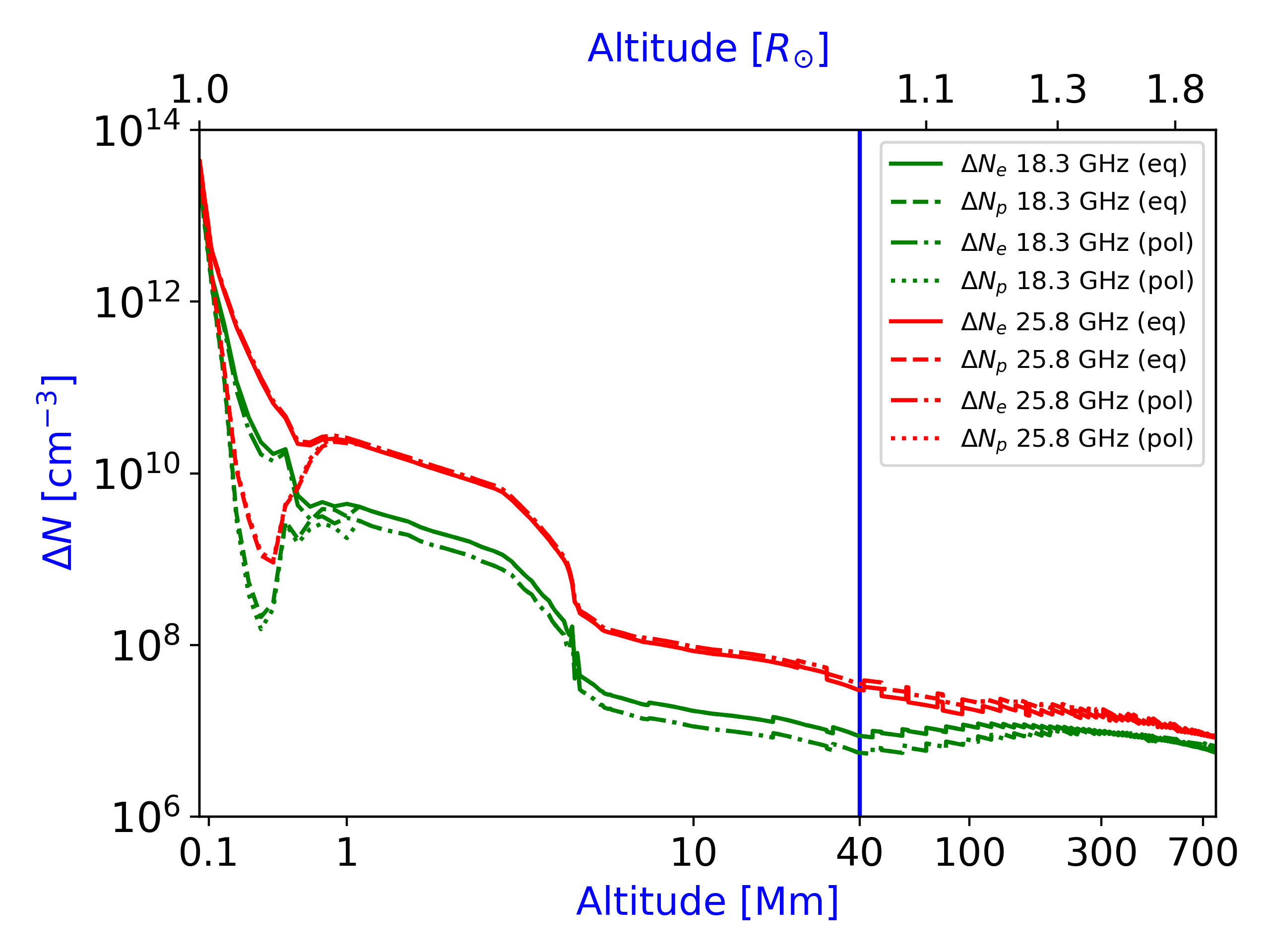}} \\
\caption{
Distribution of $N$ (electrons $N_e$ and protons $N_p$, left) and $\Delta N$ (right) as a function of the altitude from the solar surface, defined as the photosphere level $R_{\odot, opt}$.
The respective shaded area indicates the $1 \sigma$ error of these distributions.
The black line indicates the theoretical $N$ distributions of the original eSSC model.
The blue vertical line indicates the altitude $h_{SSC}$, below which the SSC model is defined.
The y-axis of these plots is shown in logarithmic scale.
Best-fit $n$ distributions are compatible (within an uncertainty $\lesssim 25\%$) with the original eSSC model below $\sim 100$~Mm; above this altitude, these distributions are characterised by high uncertainty (up to $70\%$).
}
\label{fig:profile_results_n}
\end{figure*}

As described in Sect.~\ref{par:ant_beam}, the pronounced tail signal in the observed brightness profiles shows features incompatible with hypothetical opto-mechanical deformations both for Grueff and SRT radio telescopes, suggesting a genuinely solar origin of the tail signals, rather than instrumental biases.
Furthermore, the comparison between the observed brightness profiles (averaged maps) and those obtained from the theory (SSC and eSSC models) shows that the modelled brightness profiles (Eq.~\ref{eq:t_b_model}) differ from those observed (Sect.~\ref{par:solar_radio_corona}).
Also on this occasion, these observed profiles show that the level of the averaged tails, and especially of the limb, is higher than that predicted by the theory (Fig.~\ref{fig:overview_maps_med}).
The discrepancies between the theoretical models (SSC and eSSC) and the observed brightness profiles (Fig.~\ref{fig:overview_maps_med}) led us to further investigate the behaviour of $T$, $N_e$, and $N_p$ in order to properly fit the data according to the prescriptions and limitations described in Sect.~\ref{par:solar_radio_corona}.
The obtained best-fit density and temperature distributions -- compliant with our data -- are shown in Figs.~\ref{fig:profile_results_te} and \ref{fig:profile_results_n}.
In each of these figures, the right panel indicates the difference $\Delta T$ (for the temperature) and $\Delta N$ (for the density) between the best-fit distributions and the distributions of the theoretical models (SSC and eSSC).
Thanks to our analysis, the reconstructed (and modified) $T$ and $N$ distributions of the SSC and eSSC models allow us to fit with a good agreement both the limbs and the coronal tails in our averaged $T_B$ profiles.


As we can see in Figs.~\ref{fig:profile_results_te} and \ref{fig:profile_results_n} (left panel), from our analysis we observe that the modelled $T$ and $N$ distributions are independent, within $1 \sigma$ error, on the observing frequency.
Moreover, the distributions reconstructed from the equatorial profile are compatible with the polar counterpart within $1 \sigma$, over the entire range of altitudes, and for each observing frequency.

Adopting in our analysis the SSC model, whose maximum altitude $h_{SSC}$ is indicated by the blue vertical line in Figs.~\ref{fig:profile_results_te} and \ref{fig:profile_results_n}, the reconstructed $T$ (Fig.~\ref{fig:profile_results_te}, left panel) and $N$ (Fig.~\ref{fig:profile_results_n}, left panel) distributions are compatible within $1 \sigma$ with this model.
In particular, the reconstructed $n$ distributions (Fig.~\ref{fig:profile_results_n}, left panel) are in good accordance with the literature in the radio domain (e.g. \citealp{SaintHilaire12,Ramesh20}).
A coronal analysis of the Sun \citep{ChiuderiDrago99} -- assuming hydrostatic equilibrium and using EUV and radio data -- shows an electron density derived at the base of the corona ($h \lesssim 35$~Mm) $N_e(0) \simeq 3 \times 10^8$~cm$^{-3}$, in good accordance (within $3 \sigma$) with our results ($N_e(0) \simeq (2.2 \pm 0.4) \times 10^8$~cm$^{-3}$) and the SSC model.
As shown in Fig.~\ref{fig:profile_results_te} (left panel), the $T$ distribution reaches a maximum of $(1.5 \pm 0.2) \times 10^6$~K at $h_{SSC}$, compatible with the finding of other authors in the radio domain (e.g. \citealp{Newkirk61,Gary87,Mann99,SaintHilaire12}).
On the other hand, this value is lower -- but compatible within $5 \sigma$ -- than $2.2 \times 10^6$~K obtained from other authors in the MHz range \citep{Mercier15,Vocks18}.
In the EUV domain, the $T$ distribution in the coronal region falls below $10^6$~K when coronal holes are observed in this frequency domain (e.g. \citealp{Dulk77,David98,Fludra97,Fludra99,David98}).

\begin{figure*}
\centering
{\includegraphics[width=89mm]{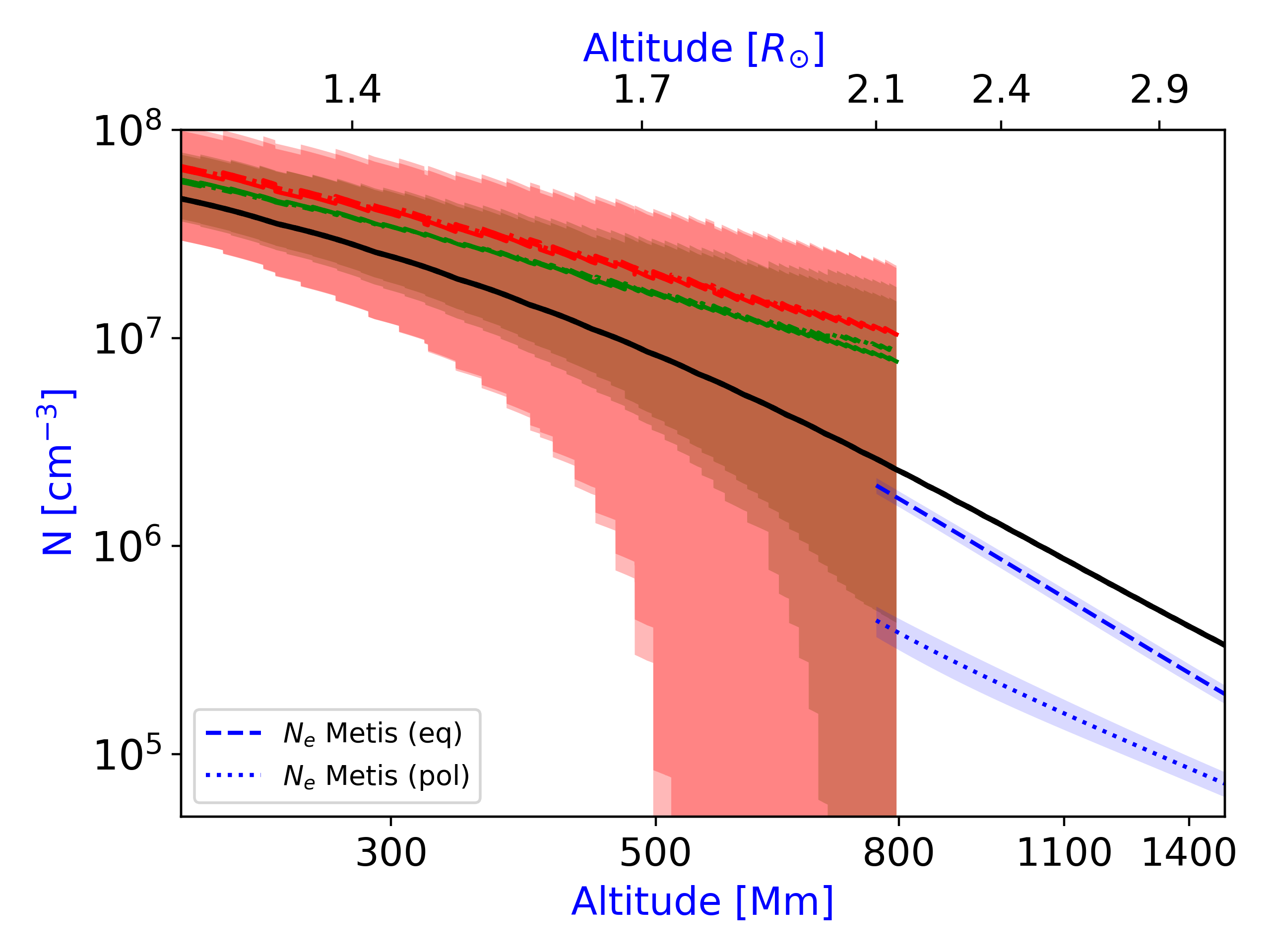}} \quad
{\includegraphics[width=89mm]{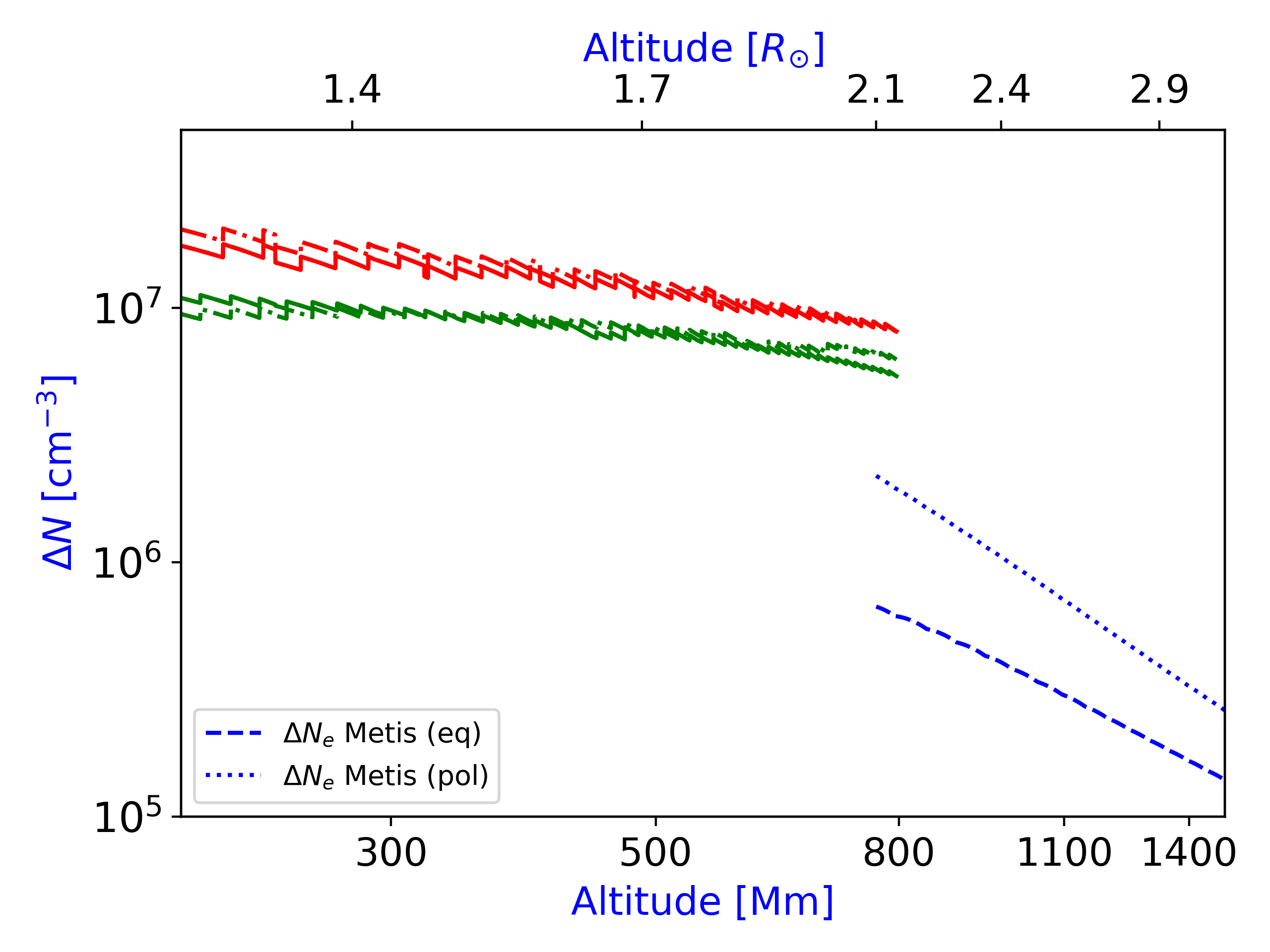}} \\
\caption{
Distribution of $N$ (left) and $\Delta N$ (right) as a function of the altitude from the solar surface.
Green and red lines indicate the measured $N_e$ obtained through observations with Grueff at $18.3$~GHz (green) and $25.8$~GHz (red) during the minimum solar activity ($2018$ -- $2020$), respectively (see the legend of the right panel in Fig.~\ref{fig:profile_results_n}).
The shaded area indicates the $1 \sigma$ error of the distributions.
Blue lines indicate the measured $N_e$ obtained through observations with Metis (dashed line: equatorial component; dotted line: polar component) during the ascending phase of the solar cycle (March $2022$).
Black line indicates the theoretical $N_e$ associated with the original eSSC model.
The y-axis of these plots is shown in logarithmic scale.
In the right panel, the difference between $\Delta N_e$ referred to Metis data is reported in terms of absolute value.
}
\label{fig:dens_sundish_metis}
\end{figure*}

Using the eSSC model in the analysis, our data can be fitted (compatibility within $1 \sigma$) up to (1) $\sim 60$~Mm of altitude ($1.09\,R_{\odot}$) for $T$, and (2) $\sim 960$~Mm of altitude ($2.4\,R_{\odot}$) for $N$.
The application of the Eq.~\ref{eq:t_b_model} -- with only $T$ as free parameter to model the pronounced tails in our $T_B$ profiles -- produces strong degeneracy and high uncertainties for altitudes higher than $\gtrsim 60$~Mm (Fig.~\ref{fig:profile_results_te}, right panel).
This suggests that the estimation of the $T$ distribution using the eSSC model is not accurate above $60$~Mm.
The modelling using $N$ ($N_e$ and $N_p$) as free parameter produces good results up to $\sim 100$~Mm ($1.14\,R_{\odot}$), where the uncertainty is $< 25\%$ (Fig.~\ref{fig:profile_results_n}).
Above this altitude, the modelled $N$ distribution decreases over altitude at a much slower rate than the distribution defined by the eSSC model, reaching an uncertainty of $\sim 70\%$ on the measurement of $N$ at $\sim 800$~Mm ($2.15\,R_{\odot}$).
Although the excess of our $N$ extrapolations at high altitudes with the eSSC model is characterised by high uncertainty, in this range of altitudes the $N$ distributions are compatible within $1\sigma$ level with the measurements obtained on March $22$ -- $25$, 2022 by the Metis coronagraph \citep{Antonucci20,Fineschi20}, aboard the Solar Orbiter ESA-NASA observatory (e.g. \citealp{Muller20}), as shown in Fig.~\ref{fig:dens_sundish_metis}.
The latter values of $N_e$ were estimated by applying the inversion procedure described in \citet{VanDeHulst50} to the Metis visible light images of level 2 (L2, data release ``V01''), as described in \citet{DeLeo23}.
We note that this comparison is affected by the fact that (1) Metis and radio observations were not simultaneous so far (coronal density depends on the solar activity), and (2) $N_e$ inferred from Metis is obtained through a different approach.

The discrepancies of our reconstructed $N$ distributions (right panel of Figs.~\ref{fig:profile_results_n} and \ref{fig:dens_sundish_metis}) -- especially in the chromospheric and the upper coronal regions -- could explain the tails observed in our $T_B$ profiles, more pronounced than those predicted by the theory (Fig.~\ref{fig:overview_maps_med}).
We pointed out that above $h_{SSC}$ (1) the $N$ distributions are characterised by an ever greater uncertainty (up to $\sim 70\%$) and by a slope lower than the distribution defined by the eSSC model (Fig.~\ref{fig:dens_sundish_metis}), and (2) the $T$ distribution is characterised by high uncertainty.
This discrepancy could be ascribed to the simple assumption of pure free-free emission in LTE conditions in the eSSC model, that allows us to model the solar atmosphere up to $\sim 2000$~Mm of altitude.
This model might lead to an overestimate of $T$ and $N$ under the hypothesis of the likely presence of additional emission components (e.g., gyro-magnetic emission; \citealp{White97}) or other effects (e.g., weak plasma emission, spicules, and refraction effects), neglected in the SSC/eSSC model (Sect.~\ref{par:solar_radio_corona}).
Moreover, this discrepancy could be explained by the fact that the worse angular resolution at low frequencies than the higher frequencies (e.g., visible, EUV, X-rays) overestimates the radio $T_B$, resulting in a higher value of $N$ in the corona (e.g. \citealp{ChiuderiDrago99}).
The assessment of this discrepancy should be addressed by multi-frequency campaigns -- such as simultaneous observations with SunDish telescope in radio and various currently-operating visible light coronagraphs such as Metis/Solar Orbiter, LASCO/SOHO \citep{Brueckner95}, and COR2/STEREO-A \citep{Howard08} -- in order to properly account for the expected coronal variability.
This campaign is crucial to improve the modelling of the solar atmosphere and to better understand the behaviour of the $T$ and $N$ distributions.
Furthermore, to disentangle these discrepancies and possible polar/equatorial anisotropies of $T_B$ (and then of the density) of the coronal radio tails, deeper exposures would be required up to two degrees from the solar centroid, improving sensitivity.

In addition to the eSSC model, in the literature other models try to describe the quiet solar atmosphere and its radio emission at different frequencies, such as the 3D numerical code called PAKAL \citep{DeLaLuz08,DeLaLuz10} and its updated version PAKALMPI \citep{DeLaLuz11}.
This model includes distributions for $T$ (extended up to $10^5$~Mm) and $N$ (extended up to $10^3$~Mm) slightly different than those both included in the eSSC model, and in our results.
In particular, the $T$ distribution of PAKAL is compatible with our $T$ distribution up to $h_{SSC}$.
For altitudes higher than $h_{SSC}$, the PAKAL distribution reaches a maximum value of $\sim 2 \times 10^6$~K at $\sim 300$~Mm, where our $T$ distribution shows strong degeneracy.
The $N$ distribution of PAKAL is compatible with our $N$ distribution up to $\sim 100$~Mm.
Above this altitude, our $N$ distribution decreases over altitude at a much slower rate than the distribution defined with PAKAL, reaching a difference of density of about one order of magnitude at $\sim 300$~Mm.

According to the discrepancy of the $T_B$ profiles in the limb regions, a preliminary modelling in the radio domain ($18$ -- $26$~GHz) using our solar maps convolved with the empirical CCBH-model (Sect.~\ref{par:ant_beam}) seems to show the presence of faint limb brightening (whose level ranges between $3 \%$ and $7 \%$), in agreement with other works (e.g. \citealp{Selhorst03}).
As stated by several authors (e.g. \citealp{Selhorst05,Menezes22}), high limb brightening observed in the solar disk can produce an overestimation in the determination of $R_{\odot}$, resulting in an underestimation of the limb brightening itself.
This underestimation could be also due to the presence of chromospheric features such as spicules located close to the limb, usually not included in the theoretical models \citep{Selhorst19b}.
Moreover, since the SSC/eSSC model has an ideal assumption of straight light of sight integral (Eq.~\ref{eq:t_b_model}) -- true for close-to-center regions -- refractive effects (e.g. \citealp{Smerd50,Alissandrakis94,Vocks18}) could deviate from a straight line the radio wave propagation (especially at MHz-level, see Fig.~\ref{fig:tb_spectra_example}) in the limb regions, causing the appearance of the limb brightening in the model.
A deeper analysis of this phenomenon is beyond the scope of this paper and it will be the subject of a future analysis.

Finally, it is worth noting that the analysis of $R_{\odot}$ through our solar maps \citep{Marongiu24a} showed that the measurements of $R_{\odot}$ -- obtained with a specific 2D model\footnote{In this model, tailored to empirically describe both the solar disk and the atmospheric emission, the solar signal is defined as the combination between the ECB-model and a 2D-Gaussian function (2GECB-model).} -- range between the values predicted by the SSC/eSSC models and the most widely used methods for estimating the radio $R_{\odot}$ in the literature (e.g. \citealp{Costa99,Selhorst11,Emilio12,Menezes17,Alissandrakis17,Menezes21}).
This aspect suggests that the methods aimed at the measurements of $R_{\odot}$, when they are improved by the addition of specific features of the solar disk and the atmosphere (such as the coronal emission level and the limb brightening), allows us to obtain values of $R_{\odot}$ closer to the canonical value $R_{\odot,opt}$ and more accurate.

\section{Summary}
\label{par:concl_svil}

In this paper we focused on the analysis of the brightness profiles of solar images obtained in the frame of the SunDish project.
We used $290$ single-dish observations throughout 5 years -- from $2018$ to mid-$2023$ -- corresponding to about half solar cycle at the radio K-band ($18$ -- $26$~GHz).
For our analysis we also used two averaged solar maps at $18.3$ and $25.8$~GHz acquired at Medicina during the minimum solar activity ($2018$ -- $2020$).
We analysed the weak tailing emission seen outside the solar disk to probe its physical origin.
In this context, we studied -- adopting a Bayesian approach -- the role of the antenna beam pattern on our solar maps, through the modelling of the solar signal (the observed brightness profiles) with a horned 2D function (CCBH model) convolved with the beam pattern.
Moreover, we constrained the $T$ and $N$ distributions of the atmospheric layers through fitting procedures of the observed brightness profiles, adopting the SSC model and its extended version (eSSC).

The results of our analysis are summarised as follows:
\begin{itemize}
    \item We probed significant emission of physical origin of the weak tailing emission seen outside the solar disk, taking into account all possible biases related to instrumental effects.
    In particular, the analysis of the role of the antenna beam pattern on our solar maps allowed us to prove the physical origin of the coronal tail emission in our images since no realistic beam pattern can produce these tails as artefacts.
    In this context, our images are important not only to improve the existing solar atmospheric models, but also to probe directly different high-lying atmospheric layers of the corona.
    \item The comparison between the observed brightness profiles (averaged maps) and those obtained from the theory (SSC and eSSC models) shows that the modelled brightness profiles differ from those observed in the coronal tail (where the observation is more pronounced than the model) and, especially, in the limb of the solar disk (where the model is more pronounced than the observation).
    \item In particular, the SSC model is in good agreement with our solar maps up to the altitude $h_{SSC}$, corresponding essentially to the base of the corona.
    Above $\sim 60$~Mm of altitude for $T$ and $\sim 100$~Mm of altitude for $N$, our results obtained with the eSSC model reveal a discrepancy with respect to the original $T$ and $N$ distributions, especially in the upper chromospheric and coronal regions.
    These discrepancies could be also ascribed to (1) the simple assumption of pure free-free emission in LTE conditions (especially in the eSSC model) and the lack of neglected effects in the eSSC model (e.g. gyro-magnetic emission), and (2) the worse angular resolution at our radio frequencies than the higher frequencies (e.g., visible, EUV, X-rays) that overestimates $N$ in the corona (e.g. \citealp{ChiuderiDrago99}).
\end{itemize}

Future observations and detailed theoretical analysis with Grueff and SRT -- for longer periods of time and with a multi-frequency approach thanks also to the new PON receivers operating up to $116$~GHz –- are crucial to probe different layers of the solar atmosphere over time.
Furthermore, simultaneous Radio-Visible Light observations (e.g. with SunDish and Metis) multi-frequency observations are necessary to properly account for the expected coronal variability, and to better understand the behaviour of the $N$ distributions, also using the same method to extrapolate the density.
A detailed theoretical analysis of the coronal model, with the analysis of the distributions of the solar medium in the context of the thermal bremsstrahlung emission mechanism plus gyro-magnetic components, is beyond the scope of this paper and it will be the subject of a future paper.
In addition, these observations are needed to better clarify several aspects, such as (1) the correlation between solar activity and the size of the Sun, (2) the polar and equatorial trends of the solar atmosphere, and (3) the question of the limb brightening (especially the presence of the polar limb brightening during the solar minima).
A preliminary investigation of the possible presence of limb brightening in our solar maps seems to show the presence of faint limb brightening (whose level ranges between $3 \%$ and $7 \%$), in agreement with other works (e.g. \citealp{Selhorst03}), but this kind of investigation is beyond the scope of this paper and it will be the subject of a future analysis.


\begin{acknowledgements}

The Medicina radio telescope is funded by the Ministry of University and Research (MUR) and is operated as National Facility by the National Institute for Astrophysics (INAF).

The Sardinia Radio Telescope is funded by the Ministry of University and Research (MUR), Italian Space Agency (ASI), and the Autonomous Region of Sardinia (RAS) and is operated as National Facility by the National Institute for Astrophysics (INAF).  

The Enhancement of the Sardinia Radio Telescope (SRT) for the study of the Universe at high radio frequencies is financially supported by the National Operative Program (Programma Operativo Nazionale - PON) of the Italian Ministry of University and Research "Research and Innovation 2014-2020", Notice D.D. 424 of 28/02/2018 for the granting of funding aimed at strengthening research infrastructures, in implementation of the Action II.1 – Project Proposal PIR01\_00010.
Solar Orbiter is a space mission of international collaboration between ESA and NASA, operated by ESA.
Metis was built and operated with funding from the Italian Space Agency (ASI), under contracts to the National Institute of Astrophysics (INAF) and industrial partners.
Metis was built with hardware contributions from Germany (Bundesministerium für Wirtschaft und Energie through DLR), from the Czech Republic (PRODEX) and from ESA.
We acknowledge the Computing Centre at INAF - Istituto di Radioastronomia for providing resources and staff support during the processing of solar data presented in this paper.
\end{acknowledgements}

\bibliographystyle{aa}            
\bibliography{aanda}          

\end{document}